\documentclass{ws-book975x65}

\usepackage{ws-book-har}           

\def\pc{\,{\rm pc}}

\def\Mpc{\,{\rm Mpc}}
\def\hmpcinv{\,{\rm h\,Mpc^{-1}}}
\def\hinvmpc{\,{\rm h^{-1}Mpc}}

\def\gcmm3{{\,{\rm g\,cm^{-3}}}}

\def\kmsMpc{\,{\rm km/s/Mpc}}

\def\ode{{\Omega_{\rm DE}}}
\def\rhode{{\rho_{\rm DE}}}

\newcommand{\fsky}{f_{\rm sky}}
\def\Msun{M_\odot}

\begin{document}

\chapter{The Accelerating Universe}
\vspace{-0.0cm} 
           
\author{Dragan Huterer \\
Department of Physics, University of Michigan, Ann Arbor, MI 48109}
\date{\today}

\section{Introduction and History} \label{sec:intro}

In this article we review the discovery of the accelerating universe
using type Ia supernovae. We then outline ways in which dark energy --
component that causes the acceleration -- is phenomenologically described. We
finally describe principal cosmological techniques to measure large-scale
properties of dark energy. This chapter therefore complements articles by
Caldwell and \citet{Linder_thisbook} in this book who describe theoretical
understanding (or lack thereof) of the cause for the accelerating universe.

\bigskip
{\bf Evidence for the missing component.}  Inflationary theory \cite{Guth80}
explains how tiny quantum-mechanical fluctuations in the early universe could
grow to become structures we see on the sky today. One of the factors that
motivated inflation is that it predicts that the total energy density relative
to the critical value is unity, $\Omega\equiv \rho/\rho_{\rm crit}=1$. This
inflationary prediction convinced many theorists that the universe is
precisely flat.

Around the same time that inflation was proposed, a variety of dynamical
probes of the large-scale structure in the universe were starting to indicate
that the {\it matter} energy density is much lower than the value needed to
make the flat. Perhaps the most specific case was made by the
measurements of the clustering of galaxies, which are sensitive to the
parameter combination $\Gamma\equiv \Omega_M h$, where $\Omega_M$ is the
energy density in matter relative to critical, and $h$ is the Hubble constant
in units of $100\kmsMpc$. The measured value at the time was $\Gamma\simeq
0.25$ (with rather large errors). One way to preserve a flat universe was to
postulate that the Hubble constant itself was much lower than the measurements
indicated ($h\sim 0.7$), so that $\Omega_M=1$ but $h\sim 0.3$
\cite{Silk_Turner}. Another possibility was the presence of Einstein's
cosmological constant (see the Caldwell article in this book), which was
suggested as far back as 1984 as the possible missing ingredient that could
alleviate tension between data and matter-only theoretical predictions
\cite{peebles84,turner84} by making the universe older, and allowing flatness
with a low value of the matter density.

\section{Type Ia supernovae and cosmology}\label{sec:SNIa}

The revolutionary discovery of the accelerating universe took place in the
late 1990s, but to understand it and its implications, we have to step back a
few decades.

\bigskip
{\bf Type Ia supernovae.}  Type Ia supernovae (SN Ia) are explosions seen to
distant corners of the universe, and are thought to be cases where a rotating
carbon-oxygen white dwarf accretes matter from a companion star, approaches
the Chandrasekhar limit, starts thermonuclear burning, and then explodes. The
Ia nomenclature refers to spectra of SN Ia, which have no hydrogen, but show a
prominent Silicon (Si II) line at 6150\AA.

SN Ia had been studied extensively by Fritz Zwicky who also gave them their
name \cite{Baade_Zwicky}, and by Walter Baade, who noted that SN Ia have very
uniform luminosities \cite{Baade_1938}. Light from type Ia supernovae
brightens and fades over a period of about a month; at its peak flux,  a SN Ia
can be a sizable fraction of the luminosity of the entire galaxy in which it
resides.

\bigskip
{\bf Standard candles.} It is very difficult to measure {\it distances} in
astronomy. It is relatively easy to measure the angular location of an object;
we can also get excellent measurement of the object's redshift $z$ from
its spectrum, by observing the shift of known spectral lines due to expansion
of the universe ($1+z=\lambda_{\rm observed}/\lambda_{\rm emitted}$). But the
distance measurements traditionally involve empirical --- and uncertain ---
methods: parallax, period-luminosity relation of Cepheids, main-sequence
fitting, surface brightness fluctuations, etc. Typically, astronomers
construct an unwieldy ``distance ladder'' to measure distance to a
galaxy: they use one of these relations (say, parallaxes -- apparent shifts
due to Earth's motion around the Sun) to calibrate distances to nearby objects
(e.g.\ variable stars Cepheids), then go from those objects to more distant
ones using another relation that works better in that distance regime. In this
process the systematic errors add up, making the distance ladder flimsy.

 ``Standard candles'' are hypothetical objects that have a nearly fixed
luminosity (that is, fixed intrinsic power that they radiate). Having standard
candles would be useful since then we could infer distances to objects just by
using the flux-luminosity inverse square law
\begin{equation}
f = {L\over 4\pi d_L^2}
\label{eq:flux_lum}
\end{equation}
where $d_L$ is the luminosity distance which can be predicted given the
object's redshift and contents of the universe (i.e.\ energy densities of matter and
radiation relative to the critical density which makes the universe spatially
flat, as well as other components such as radiation).  In fact, we don't even
need to know the luminosity of the standard candle to be able to infer {\it
  relative} distances to objects.

In astronomy, flux is often expressed in terms of apparent magnitude -- a logarithmic measure of
flux, and luminosity is related to the absolute magnitude of the object. So,
in astronomical units, Eq.~(\ref{eq:flux_lum}) reads
\begin{equation}
 m-M = 5 \log_{10}\left ({d_L\over 10\pc}\right )
\label{eq:DM}
\end{equation}
where the quantity on the left-hand side is also known as the {\it distance
  modulus}. For an object that is 10 parsecs away, the distance modulus is
zero. For a standard candle, the absolute magnitude $M$ (or, equivalently,
luminosity $L$) is known to be approximately the same for each
object. Therefore, measurements of the apparent magnitude to each object
provide information about the luminosity distance, and thus the makeup of the
universe.  

\bigskip
{\bf Finding SN.} The fact that SN Ia can potentially be used as a standard
candle has been realized long ago, at least as far back as the 1970s
\cite{Kowal,Colgate}. However, a major problem is to find a method to schedule
telescopes to discover SN before they happen.  If we point a telescope at a
galaxy and wait for the SN to go off, we will wait several hundred
years. There had been a program in the 1980s to find supernovae
\cite{Norgaard_Nielsen} but, partly due to inadequate technology and equipment
available at the time, it discovered only one SN, and after the peak of the
light-curve.

The first major breakthrough came in the 1990s when two teams of SN
researchers Supernova Cosmology Project (SCP; led by Saul Perlmutter and
organized in the late 1980s) and High-z Supernova Search Team (Highz;
organized in the mid 1990s and led, at the time, by Brian Schmidt) developed
an efficient approach to use world's most powerful telescopes working in
concert to discover and follow up high-redshift SN, and thus complement the
existing efforts at lower redshift led by the Cal\'{a}n/Tololo collaboration
\cite{Calan-Tololo}.  These teams had been able to essentially guarantee
that they would find batches of SN in each run. [For popular reviews of these
  exciting developments, see \citet{Kirshner_book} and
  \citet{Perlmutter_Schmidt}.]

\begin{figure}[!t]
\begin{center}

\includegraphics[height=3.4in,width=2.1in]{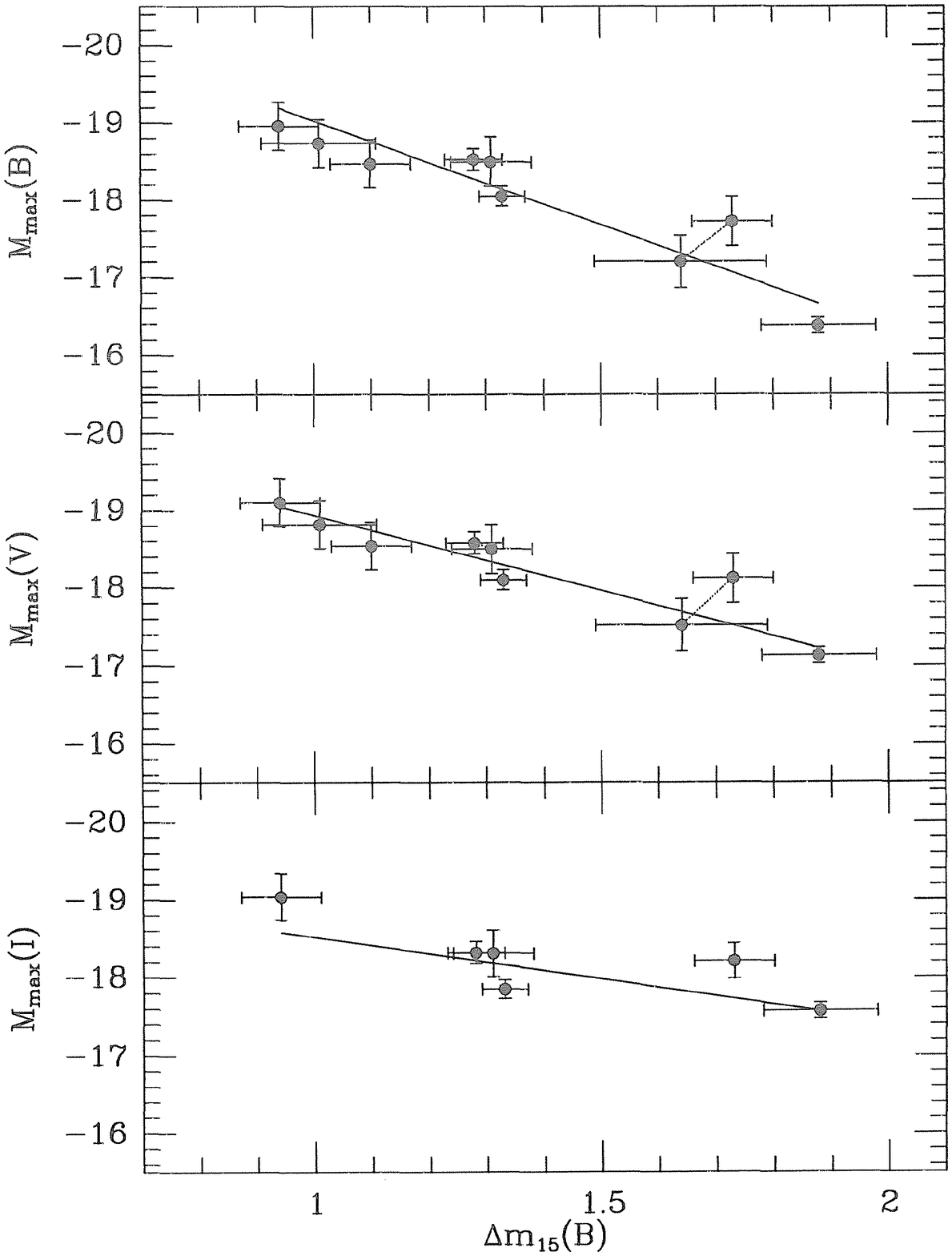}
\includegraphics[height=3.6in,width=2.3in]{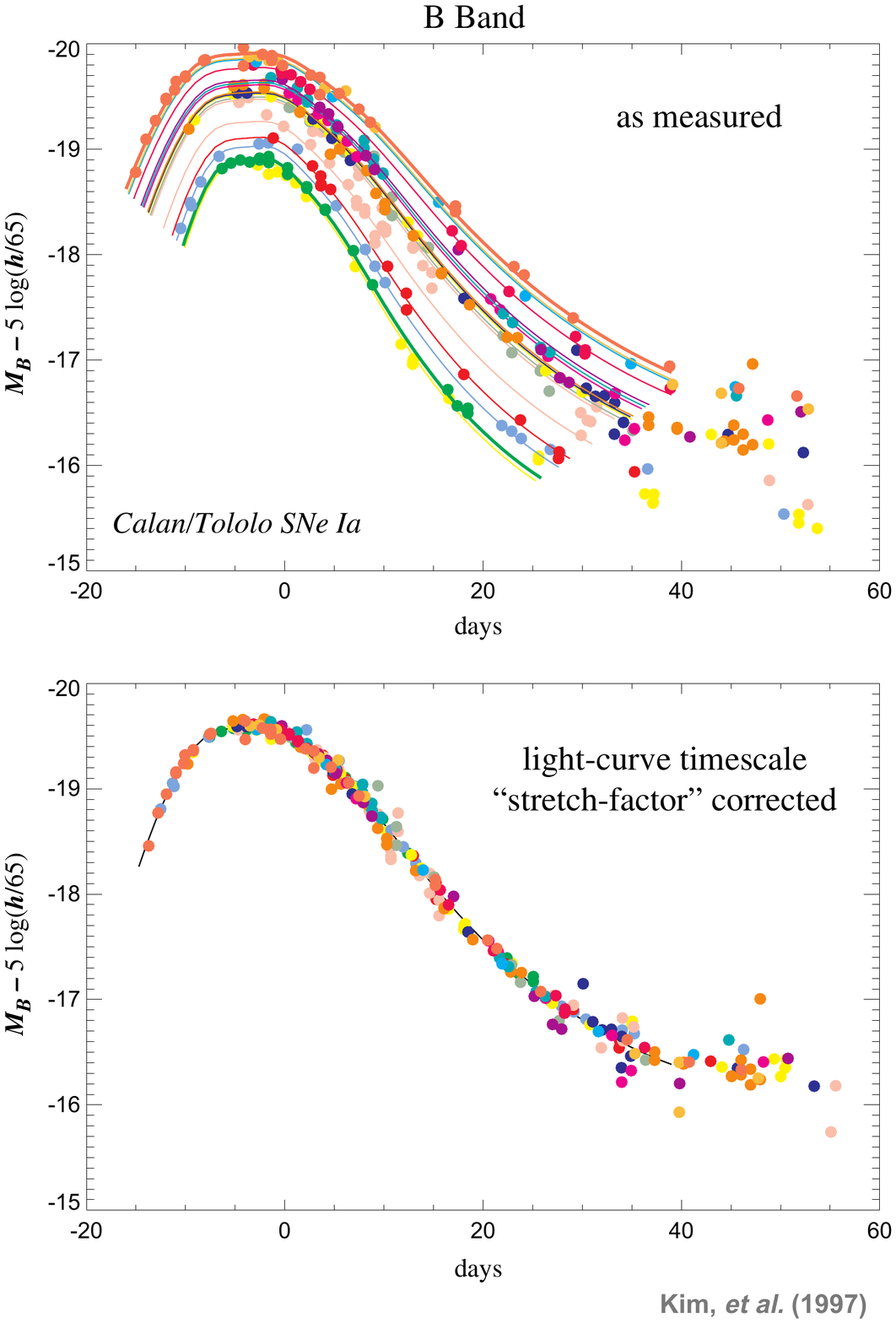}
\end{center}
\vspace{-0.7cm}
\caption{{\it Left panel:} Phillips relation, from his 1993 paper. The
  (apparent) magnitude of type Ia supernovae is correlated with $\Delta
  m_{15}$, the decay of the light curve 15 days after the maximum. {\it Right
    panel: } light curves of a sample of SN Ia before correction for stretch
  (essentially, the Phillips relation; top), and after (bottom); adopted from
  \citet{Kim97}.  }
\label{fig:Phillips_rel}
\end{figure}

The second breakthrough came in 1993 by Mark Phillips, astronomer working
in Chile \cite{Phillips_93}. He noticed that the SN luminosity -- or absolute
magnitude -- is correlated with the decay time of SN light curve. Phillips
considered the quantity $\Delta m_{15}$, 
the attenuation of the flux of SN between the light maximum and 15 days past
the maximum.  He found that $\Delta m_{15}$ is strongly correlated with the
intrinsic brightness of SN; see the left panel of
Fig.~\ref{fig:Phillips_rel}. The ``Phillips relation'' roughly goes as
\begin{center}
{\rm Broader\,\, is\,\, brighter.}
\end{center}
In other words, supernovae with broader light-curves have a larger intrinsic
luminosity. One way to quantify this relation is to use a ``stretch'' factor
which is a (calibration) parameter that measures width of a light curve
\cite{perlmutter99}; see the right panel of Fig.~\ref{fig:Phillips_rel}.  By
applying the correction based upon the Phillips relation, astronomers found
that the intrinsic dispersion of SN, which is of order $\sim 0.5$ magnitudes,
can be brought down to $\delta m \sim 0.2$ magnitudes once we correct each SN
luminosity using its stretch factor. Note that the final dispersion in
magnitudes corresponds to the error in distance of $\delta d_L/d_L=
(\ln(10)/5)\,\delta m\simeq 0.5\,\delta m\sim 0.1$.  The Phillips relation was
the second key ingredient that enabled SN Ia to achieve precision needed to
probe contents of the universe accurately.

The third key invention was the development of techniques to correct SN
magnitudes for dimming by dust, or 'extinction', out of multi-color
observation of SN light \cite{MLCS_2,MLCS_1}. Such corrections
are an important part of SN cosmology to this day \cite{Jha_MLCS2k2,SALT2,SIFTO}.

Finally, the fourth and perhaps most important ingredient for the discovery of dark
energy was development and application of charge-coupled devices (CCDs) in
observational astronomy. Both teams of SN hunters used the CCDs, which had originally
been installed at telescopes at Kitt Peak and Cerro Tololo
\cite{Kirshner_review}.

Some of the early results came out in the period of 1995-1997; however  these
results were based on a handful of high-redshift SN and had large errors
(e.g.\ \cite{Perlmutter_97,Garnavich_97,Perlmutter_SN_halfage}).

\begin{figure}[tb]
\begin{center}
\includegraphics[height=2.8in,width=4.1in]{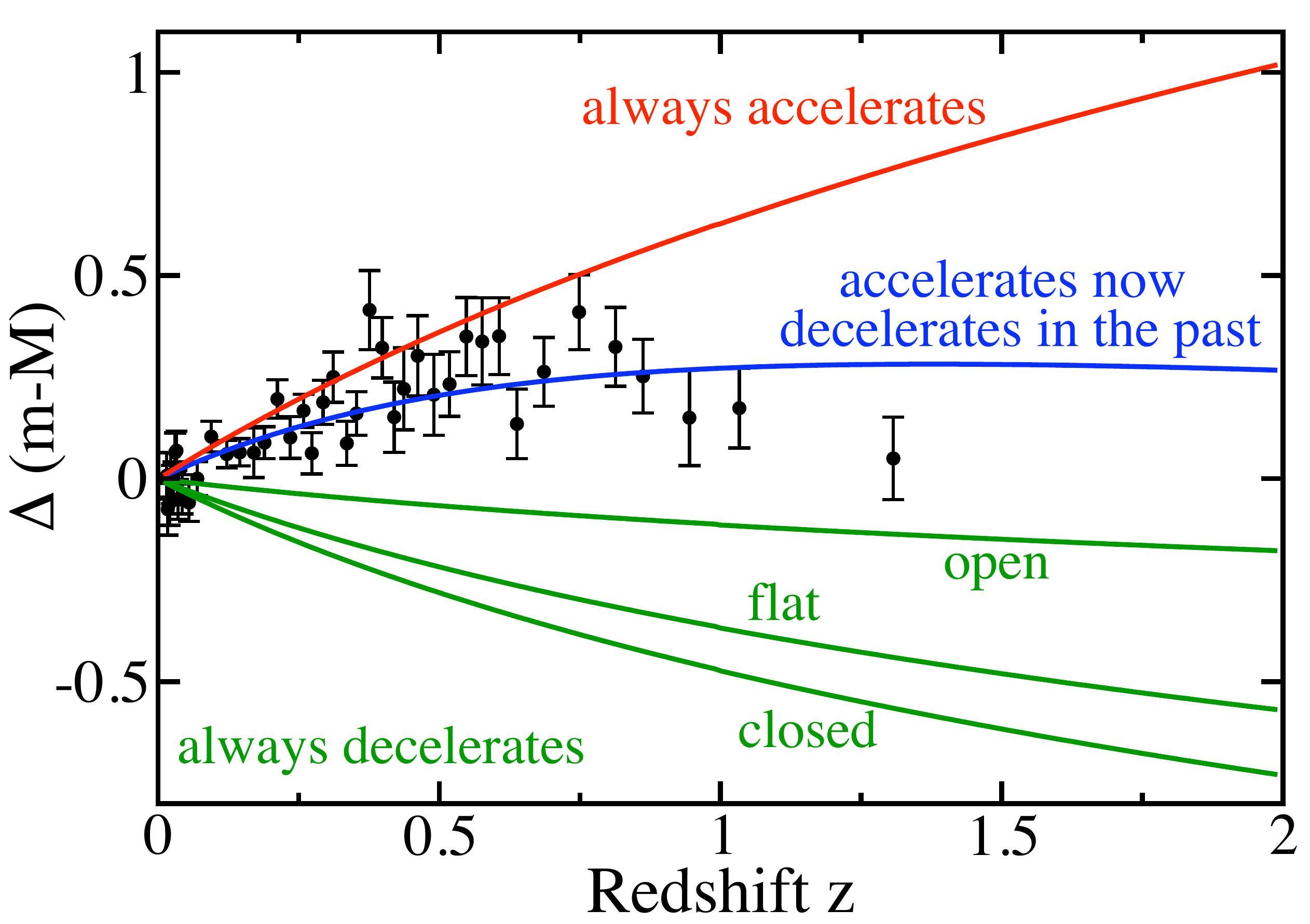}
\end{center}
\caption{Evidence for transition from deceleration in the past to acceleration
  today.  The blue line shows a model that fits the data, where acceleration
  happens at late epochs in the history of the universe (i.e.\ starting a few
  billion years ago, and billions of years after the Big Bang). For
  illustration, we also show three representative matter-only models in green,
  with open, closed and flat geometry. Finally, the red curve shows a model
  that always exhibits acceleration, and it too does not fit the SN data which
  show a characteristic ``turnover'' in the magnitude vs.\ redshift plot.  The
  plot uses binned data from the Union2 compilation \cite{Union2} containing
  557 SN.  }
\label{fig:deceleration}
\end{figure}

\bigskip
{\bf The discovery of dark energy.}  The definitive results, based on $\sim
50$ SN by either team that combined the nearby sample previously observed by
the Cal\'{a}n/Tololo collaboration and the newly acquired and crucial sample of
high-redshift SN, came out soon thereafter \cite{riess98,perlmutter99}.  The
results of the two teams agreed, and indicated that more distant SN are dimmer
than would be expected in a matter-only universe; see
Fig.~\ref{fig:deceleration}. In other words, the universe's expansion rate is
speeding up, contrary to expectation from the matter-dominated universe with
{\it any} amount of matter and regardless of curvature.

Phrased yet differently, the data indicate universe that is
accelerating -- that is presence of a new component with strongly negative pressure.
This can easily be seen from the   {\it acceleration equation}, which is one
of Einstein's equations applied to the case of the homogeneous universe
\begin{equation}
{\ddot{a}\over a} = -{4\pi G\over 3} (\rho + 3p)
= -{4\pi G\over 3} (\rho_M + \rho_{\rm DE} +  3p_{\rm DE})
\label{eq:Friedmann_2}
\end{equation}
where $\rho$ and $p$ are the energy density and pressure of components in the
universe, assuming they are matter and a new component we call dark energy
(radiation is negligible relative to matter at redshifts much less than $\sim
10^3$, and the pressure of matter is always negligible). If the universe is
accelerating, then $\ddot{a}>0$, and the only way it can be is if the {\it
  pressure of the new component is strongly negative}. Phrased in terms of
equation of state, $w\equiv p_{\rm DE}/\rho_{\rm DE}<-1/3$ regardless of the
density of matter $\rho_M$.

The discovery of the accelerating universe with supernovae was a watershed
event in modern cosmology, and the aforementioned two discovery papers are among the
most cited physics papers of all time.  This component that makes the universe
accelerate was soon named ``dark energy'' by the theoretical cosmologist
Michael Turner \cite{reconstr}.

The SN data are illustrated in Fig.~\ref{fig:deceleration}, where upwards of
500 SN measurements from the Union2 compilation \cite{Union2} have been binned
in redshift. The blue line shows a model that fits the data, where
acceleration happens at late epochs in the history of the universe
(i.e.\ starting a few billion years ago, and billions of years after the Big
Bang). For illustration, we also show three representative matter-only models
in green, with open, closed and flat geometry, neither of which fits the data
well. Finally, the red curve shows a model that always exhibits acceleration,
and it too does not fit the SN data which show a characteristic ``turnover''
in the magnitude vs.\ redshift plot.

\bigskip
{\bf Observable and inferred quantities with SN Ia.}
The luminosity distance $d_L$ is related to the cosmological parameters via
\begin{equation}
d_L =  (1+z){H_0^{-1}\over \sqrt{\Omega_K}}
\sinh\left [ {\sqrt{\Omega_K}
\int_0^z {dz'\over \sqrt{\Omega_M (1+z)^3 + \Omega_{\rm DE} (1+z)^{3(1+w)} +
    \Omega_R(1+z)^4 + \Omega_K (1+z)^2}}
}
\right ]
\label{eq:dL}
\end{equation}
where the key term in this expression featuring $\sinh(x)$ for $\Omega_K>0$
(open universe) effectively turns into $\sin(x)$ (closed universe;
$\Omega_K<0$) or just $x$ (flat universe; $\Omega_K =0$). Here $\Omega_M$,
$\Omega_R$, and $\Omega_{\rm DE}$ are the energy densities of matter (visible
plus dark), radiation (mainly cosmic microwave background (CMB) photons), and
dark energy relative to critical density, and $\Omega_K =
1-\Omega_M-\Omega_R-\Omega_{\rm DE}$.

Now Eq.~(\ref{eq:DM}) can be rewritten as
\begin{equation}
m\equiv 5\log_{10}(H_0d_L) + \mathcal{M} 
\label{eq:mag_redshift}
\end{equation}
where the "script-M" factor is defined as
\begin{equation}
\mathcal{M} \equiv M - 5\log_{10}\left ({H_0\over \Mpc^{-1}}\right) + 25.
\end{equation}
Note that $\mathcal{M}$ is a dummy parameter that captures {\it two} uncertain
quantities: the absolute magnitude (i.e.\ intrinsic luminosity) of a
supernova, $M$, and the Hubble constant $H_0$. We typically do not know
$\mathcal{M}$, and we need to marginalize (i.e.\ integrate) over all values of
this parameter in the cosmological analysis.

The situation is now clear: astronomers measure $m$, which is inferred, for
example, from the flux at the peak of the light curve. Then they measure the
redshift of SN host galaxy. With the sufficient number of SN measurements, they
can marginalize over the parameter $\mathcal{M}$ and be left with,
effectively, measurements of luminosity distance vs.\ redshift. A plot of
either $m(z)$ or $d_L(z)$ is called the Hubble diagram.

\begin{figure}[!t]
\begin{center}
\includegraphics[height=3.0in,width=2.5in]{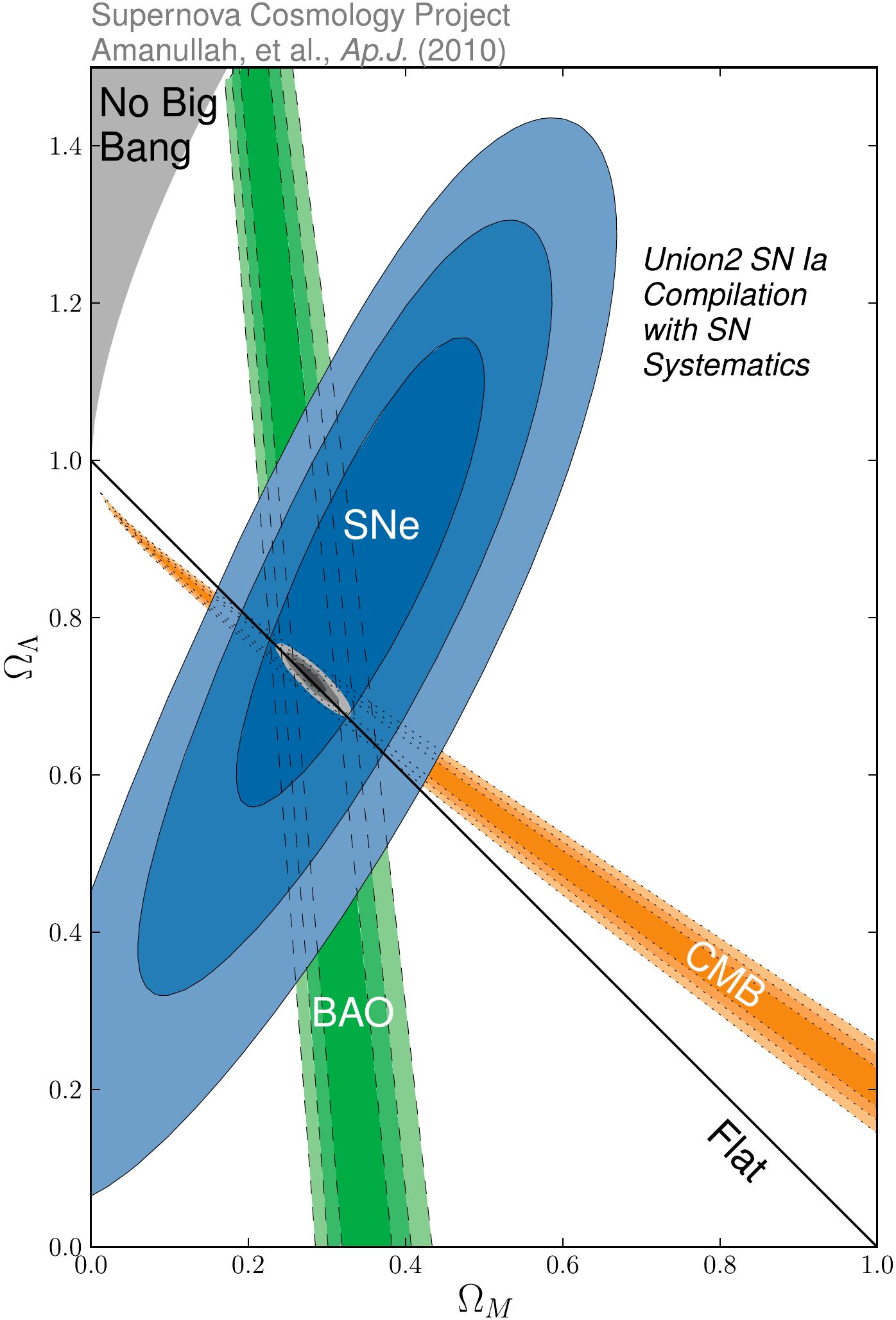}
\includegraphics[height=2.3in,width=2.3in]{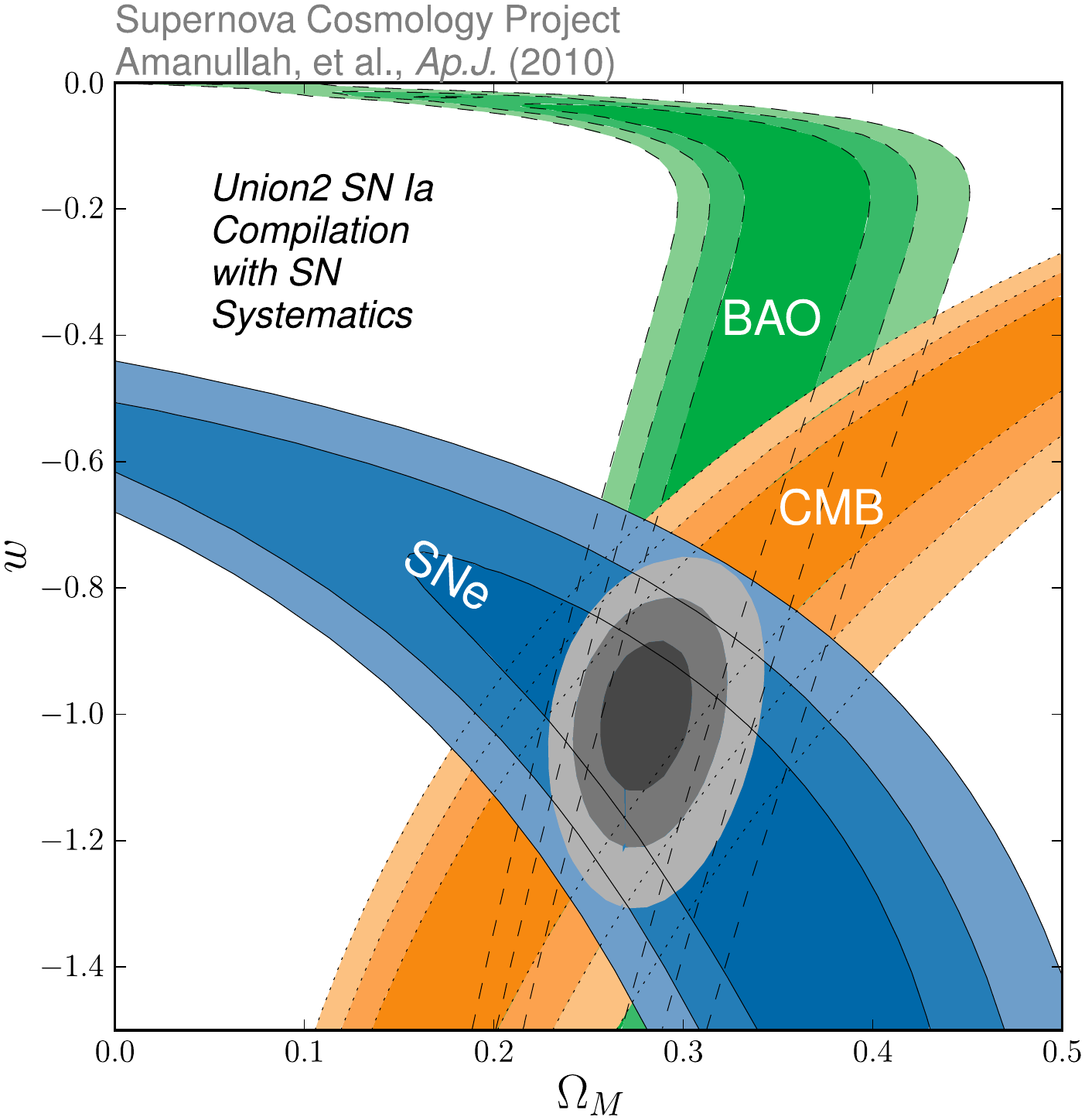}
\end{center}
\caption{{\it Left panel}: Constraints upon $\Omega_{\rm M}$ and
  $\Omega_\Lambda$ in the consensus model using baryon acoustic oscillation
  (BAO), CMB, and SN measurements.  {\it Right panel}: Constraints upon
  $\Omega_{\rm M}$ and constant $w$ in the fiducial dark energy model using
  the same data sets. From \citet{Union2}.}
\label{fig:kowalski}
\end{figure}

These results have been greatly strengthened since, with many hundreds of SN
Ia currently indicating same results, but with smaller errors, compared to the
original 1998-99 papers
\cite{Knop_03,Riess_04,SNLS,Riess_07,WoodVasey_07,SDSS_SN,hicken,Union2}. Meanwhile, other
cosmological probes have come in with results confirming the SN results (see
the right panel of Fig.~\ref{fig:kowalski}).

\bigskip
{\bf Systematic errors.} Systematic errors may creep up in SN observations,
and stand in the way of making SN Ia a more precise tool of cosmology. Here we
list a few prominent sources of error, and ways in which they are controlled:

\begin{figure}[tb]
\begin{center}
\includegraphics[height=2.4in,width=3.2in]{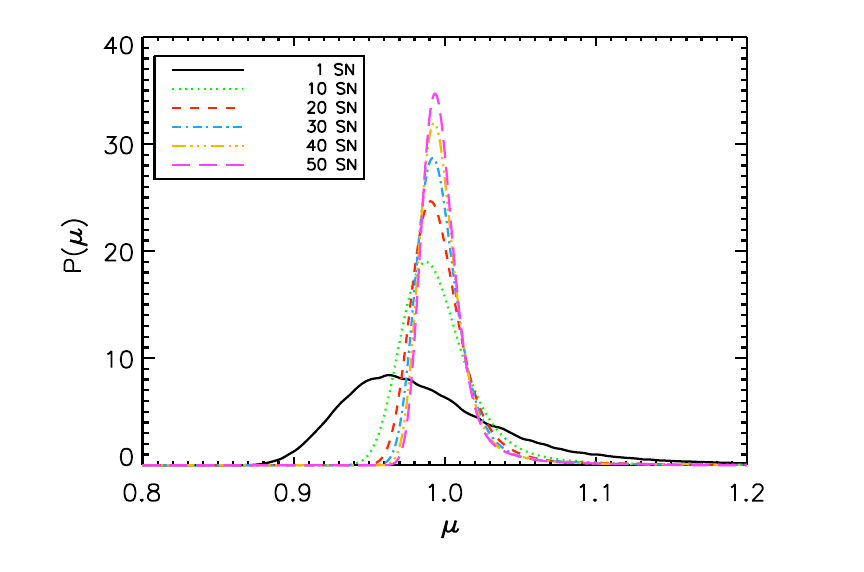}
\end{center}
\vspace{-0.7cm}
\caption{Magnification distribution for lensing of a supernova at $z=1.5$ in
  the usual $\Lambda$CDM cosmology (black curve). Other curves show how the
  distribution both narrows and becomes more gaussian as we average over more
  SN. Adopted from \citet{Holz_Linder}.
}
\label{fig:SN_lensing}
\end{figure}

\begin{itemize}

\item Extinction: is it possible that SN appear dimmer simply because of
  extinction by dust particles scattered between us and distant SN?
  Fortunately there are ways to stringently control (and correct for)
  extinction, by observing SN in different wavelength bands. But also, if
  extinction were to be responsible for the appearance of dimming, then we
  would expect more distant SN to appear uniformly more dim. Moreover, a
  ``turnover'' in the SN Hubble diagram has been clearly observed
  (e.g.\ \citet{Riess_04}) indicating that the universe is matter
  dominated at high $z$. The turnover cannot easily be explained by
  extinction.

\item Evolution: is it possible that SN evolve, so that we are seeing a
  different population at higher redshift that  is intrinsically dimmer
  (violating the assumption of a standard candle)?  SN Ia do not own a
  ``cosmic clock''; rather, they respond to their local environment, in
  addition to being ruled by the physics of accretion/explosion. So, by observing
  various signatures, in particular in SN spectra, researchers can identify
  local environmental conditions, and even go so far to compare only
  like-to-like SN (resulting, potentially, in several Hubble diagrams, one
  for each subspecies). First such comparisons have been made recently.

\item Typing: is it possible that non-Ia supernovae have crept in the samples
  used for dark energy analysis? This question is rather easy to answer, as SN
  Ia possess characteristic spectral lines which uniquely identify these
  SN. Accurate typing, however, becomes more challenging for SN surveys which cannot
  afford to take spectra of all SN; upcoming and future imaging surveys such
  as the Dark Energy Survey (DES) or Large Synoptic Survey Telescope (LSST)
  are examples. For those surveys, one will have to apply sophisticated tests
  based on photometric information alone to establish whether or not a given
  supernova is type Ia.

\item K-corrections: As SN Ia are observed at larger and larger redshifts, their
  light is shifted to longer wavelengths. Since astronomical observations are
  normally made in fixed band passes on Earth, corrections need to be applied
  to account for the differences caused by the spectrum shifting within these
  band passes, and error in these corrections needs to be tightly controlled. 

\item Gravitational lensing: distant SN are gravitationally lensed by matter
  along the line of sight, making them magnified or demagnified, and thus
  appearing brighter or dimmer. The lensing effect goes roughly as $z^2$ and
  is non-negligible only for high-z SN; $z\gtrsim 1.2$. The {\it mean}
  magnification is zero (owing to a theorem that the total light is
  conserved), but the distribution is skewed, meaning that most SN get
  demagnified but occasional ones get strongly magnified. The way to protect
  against biases due to gravitational lensing is to seek ``safety in numbers''
  \cite{Holz_Linder}: simply put, if we collect enough SN at any given
  redshift (in practice, $\sim 50$ SN per $\Delta z=0.1$), the effects of
  gravitational lensing will average down to near zero; see
  Fig.~\ref{fig:SN_lensing}.
\end{itemize}

At the present time the SN systematic errors are well controlled, and are
comparable to the statistical errors. The factor that gives undisputed
credence to the result that the universe is accelerating, however, is
confirmation with the galaxy clustering (the baryon acoustic oscillations, to
be described later in this article), and the CMB constraints; see
Fig.~\ref{fig:kowalski}. In fact, even if one completely drops the SN
constraints from the analysis, the combination of the galaxy clustering and
the CMB firmly points to the existence of dark energy!

\section{Parametrizations of dark energy}\label{sec:parametr}

\bigskip
{\bf Introduction.}  The absence of a consensus model for cosmic acceleration
presents a challenge in trying to connect theory with observations.  For dark
energy, the equation-of-state parameter $w$ provides a useful phenomenological
description; because it is the ratio of pressure to energy density, it is also
closely connected to the underlying physics.  On the practical side,
determining a free function is more difficult than measuring parameters.  We
now review a variety of formalisms that have been used to describe and
constrain dark energy.

First, let us recall some basics. From continuity equation, $\dot{\rho} + 3H(p+\rho)=0$,
we can calculate the dark energy density as a function of redshift for an
arbitrary equation of state $w(z)$
\begin{equation}
{\rhode(z)\over \rhode_{,0}} = \exp\left (3\int_0^z (1+w(z'))d\ln (1+z')\right ).
\end{equation}

\bigskip
{\bf Parametrizations.}
The simplest parameterization of dark energy is 
\begin{equation}
w={\rm const}.
\end{equation}
This form fully describes vacuum energy ($w=-1$) or topological defects
($w=-N/3$ with $N$ an integer dimension of the defect -- 0 for monopoles, 1
for strings, 2 for domain walls). Together with $\ode$ and $\Omega_{\rm
  M}$, $w$ provides a 3-parameter description of the dark-energy sector (2
parameters if flatness is assumed).  However, it does not describe scalar
field or modified gravity models which generically have a time-varying $w$.

A number of two-parameter descriptions of $w$ have been explored in the
literature, e.g., $w(z)  =  w_0 + w' z$ \cite{Cooray_Huterer}.  For low
redshift they are all essentially equivalent, but for large $z$, some lead to
unrealistic behavior, e.g., $w\ll -1$ or $\gg 1$.  The parametrization \cite{Linder_wa}
\begin{equation}
w(a)=w_0 + w_a(1-a) = w_0 + w_a \,{z\over 1+z}
\label{eq:w0wa}
\end{equation}
where $a=1/(1+z)$ is the scale factor, avoids this problem, fits many scalar
field and some modified gravity behaviors, and leads to the most commonly used
description of dark energy, namely $(\ode,\Omega_{\rm M},w_0,w_a)$. The energy
density is then
\begin{equation}
{\rhode(a)\over \rhode_{,0}}  = a^{-3(1+w_0+w_a)}e^{-3(1-a)w_a}.
\end{equation}

More general expressions have been proposed.  However one problem with
introducing more parameters is that additional parameters make the equation of
state very difficult to measure, while the parametrizations are still {\it ad
  hoc} and not well motivated from either theory or measurements' point of
view.

\begin{figure}[!t]
\begin{center}
\includegraphics[height=2.2in,width=2.7in]{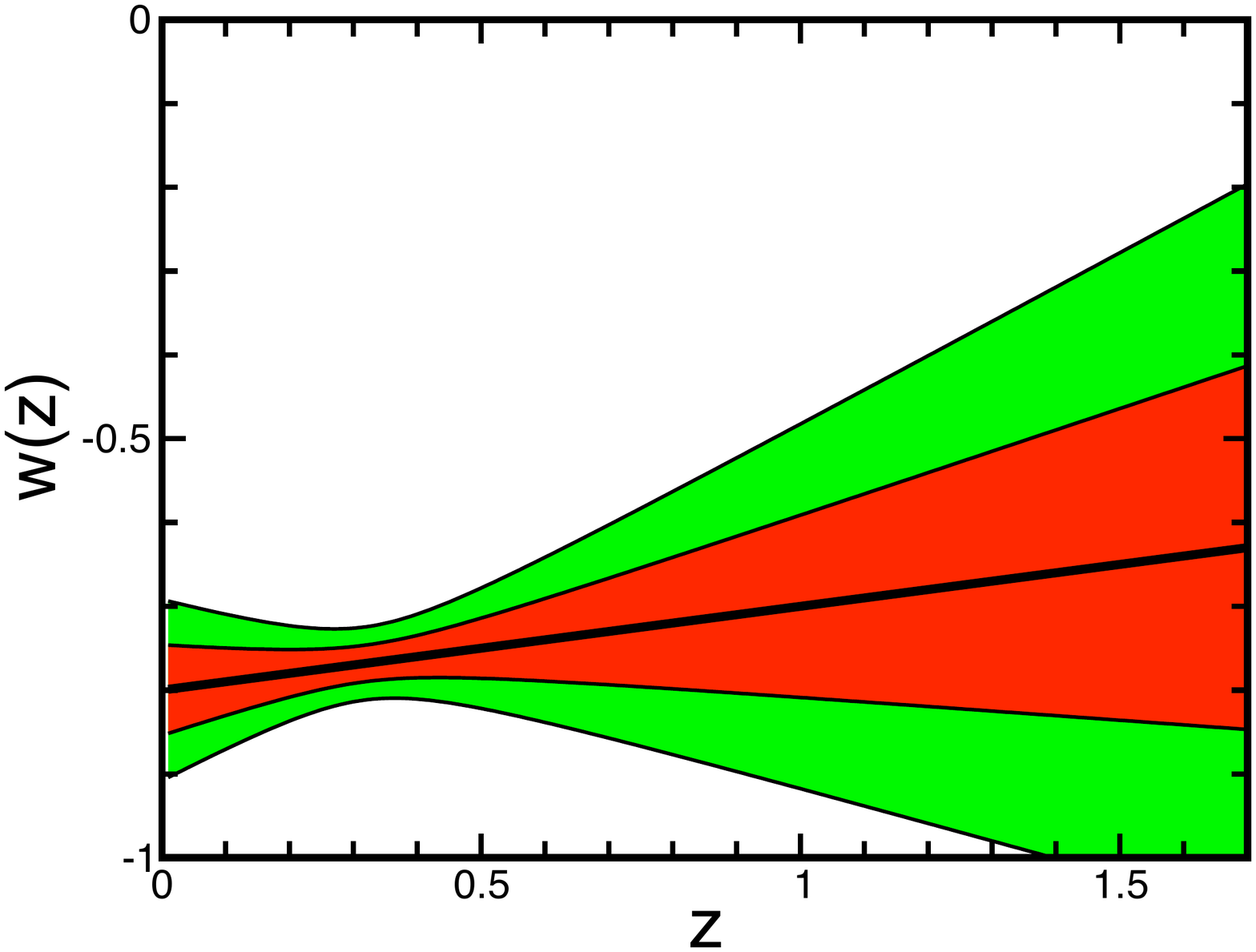}\hspace{-0.7cm}
\includegraphics[height=2.0in,width=2.4in]{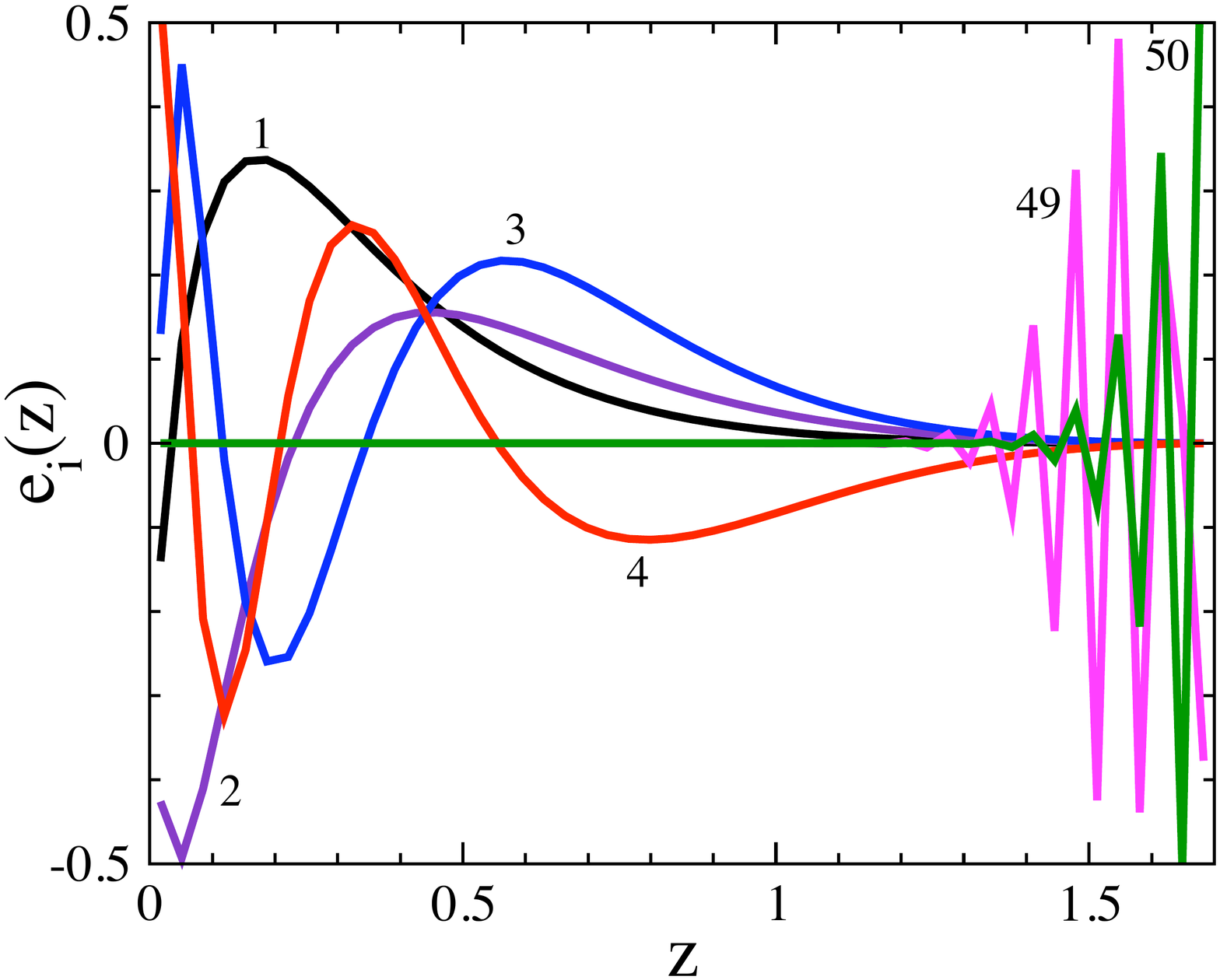}
\end{center}
\caption{{\it Left panel:} Example of forecast constraints on $w(z)$, assuming
  $w(z)=w_0+w'z$.  The ``pivot'' redshift, $z_p\simeq 0.3$, is where $w(z)$ is
  best determined. Adopted from \citet{Hut_Tur_00}. {\it Right panel:} The
  four best-determined (labelled $1$-$4$) and two worst-determined (labelled
  $49, 50$) principal components of $w(z)$ for a future SN Ia survey such as
  SNAP, with several thousand SN in the redshift range $z=0$ to
  $z=1.7$. Adopted from \citet{Huterer_Starkman}.}
\label{fig:w0wp_PC}
\end{figure}

Finally, it is useful to mention one simple way to elucidate redshift where
the measurement accuracy of the equation of state, for a given survey is
highest.  Two-parameter descriptions of $w(z)$ that are linear in the
parameters entail the existence of a ``pivot'' redshift $z_p$ at which the
measurements of the two parameters are uncorrelated and the error in $w_p
\equiv w(z_p)$ reaches a minimum; see the left panel of
Fig.~\ref{fig:w0wp_PC}. Writing the equation of state in Eq.~(\ref{eq:w0wa})
in the form
\begin{equation}
w(a) = w_p + (a_p-a) w_a
\end{equation}
it is easy to translate constraints from the ($w_0, w_a$) to ($w_p, w_a$)
parametrization, as well as determine $a_p$ (or $z_p$), for any particular
data set. This is useful, as measurements of the equation of state at the
pivot point might provide most useful information in ruling out models
(e.g.\ ruling out $w=-1$).

\bigskip
{\bf Direct reconstruction.}
Another approach is to directly invert the redshift-distance relation $r(z)$
measured from SN data to obtain the redshift dependence of $w(z)$ in terms of
the first and second derivatives of the comoving distance \cite{reconstr,
  Nakamura_Chiba, Starobinsky},
\begin{equation}
1+w(z) = {1+z\over 3}\, {3H_0^2\Omega_{\rm M}(1+z)^2 + 2(d^2r/dz^2)/(dr/dz)^3\over
        H_0^2\Omega_{\rm M}(1+z)^3-(dr/dz)^{-2}}~.
\label{eq:wz_reconstr}
\end{equation}
Assuming that dark energy is due to a single rolling scalar field, the scalar
potential $V(\phi)$ can also be reconstructed.
Others have suggested reconstructing the dark energy density
\cite{Wang_Mukherjee_03, Wang_Teg_uncorr}
\begin{equation}
\rho_{\rm DE}(z) = {3\over 8\pi G} \left[{1\over (dr/dz)^2} - \Omega_{\rm M}H_0^2 (1+z)^3 \right] \, .
\end{equation}

Direct reconstruction is the only approach that is truly
model-independent. However, it comes at a price -- taking derivatives of noisy
data.  In practice, one must fit the distance data with a smooth function, and the
fitting process introduces systematic biases.  While a variety of methods have
been pursued \cite{Hut_Tur_00,Weller_Albrecht}, it appears that direct
reconstruction is too challenging and not robust even with SN Ia data of
excellent quality (though see \citet{Holsclaw}).  And while the reconstruction
of $\rho_{\rm DE}(z)$ is easier since it involves only first derivatives of
distance, $w(z)$ is more useful a quantity since it contains more information
about the nature of dark energy than $\rho_{\rm DE}(z)$.  [For a review of
  dark energy reconstruction methods, see \citet{Sahni_review}.]

\bigskip
{\bf Principal components.}
The cosmological function that we are trying to determine --- $w(z)$,
$\rho_{\rm DE}(z)$, or $H(z)$ --- can be expanded in terms of principal
components, a set of functions that are uncorrelated and orthogonal by
construction \cite{Huterer_Starkman}. In this approach, the data determine
which components are measured best.

\begin{figure}[t]
\begin{center}
\includegraphics[height=1.8in,width=4.8in]{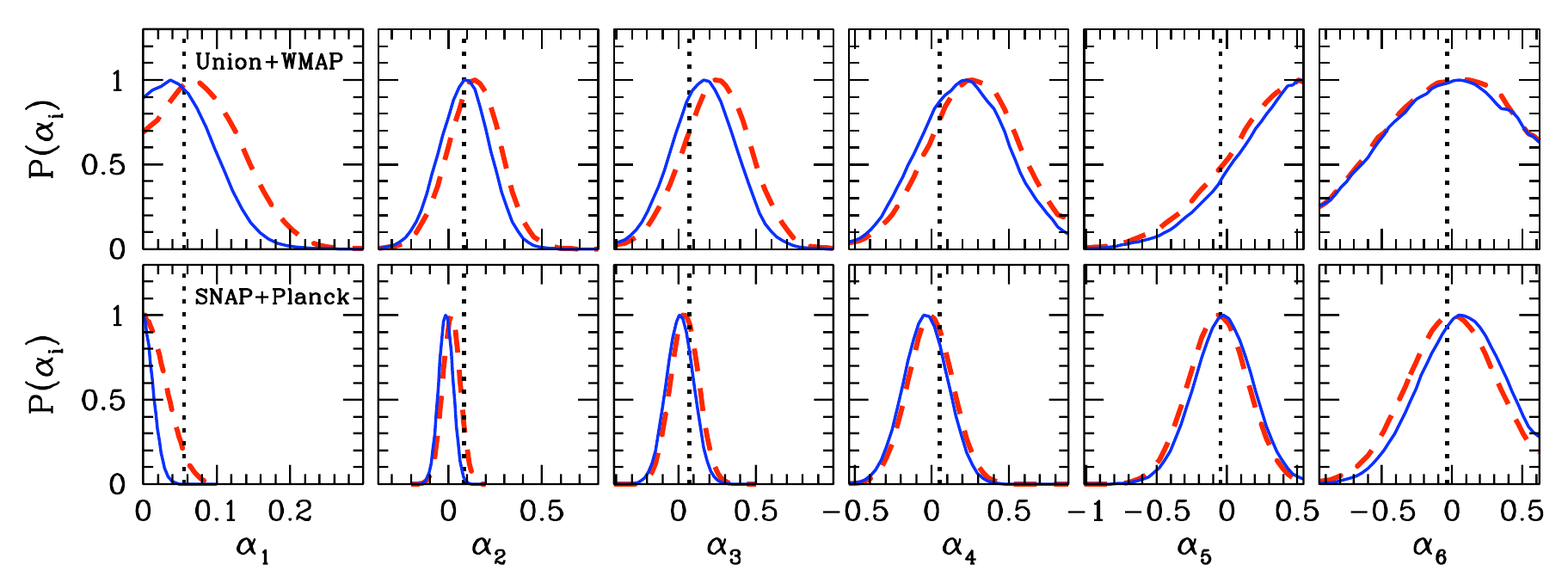}
\end{center}
\caption{ Marginalized 1D posterior distributions for the first 6 PCs of flat
  (solid blue curves) and nonflat (dashed red curves) quintessence models.
  Top row: {\it current} Union+WMAP data; note that all PCs are consistent
  with $\alpha_i=0$ (that is, $w(z)=-1$) except perhaps the fifth one. Bottom
  row: forecasts for {\it future} SNAP+Planck {\it assuming} a realization of
  the data with $\alpha_i=0$.  Vertical dotted lines show the predictions of
  an example quintessence model.  Adopted from \citet{MHH_FoM}.  }
\label{fig:pc_1d}
\end{figure}

For example, suppose we parametrize $w(z)$
in terms of piecewise constant values $w_i$ ($i=1, \ldots, N$), each defined
over a small redshift range ($z_i$, $z_i+\Delta z$).  In the limit of small
$\Delta z$ this recovers the shape of an arbitrary dark energy history (in
practice, $N\gtrsim 20$ is sufficient), but the estimates of the 
$w_i$ from a given dark energy probe will be very noisy. 
Principal Component Analysis extracts from those noisy estimates 
the best-measured features of $w(z)$. We find the eigenvectors $e_i(z)$ of
the inverse covariance matrix for the parameters $w_i$ and the corresponding
eigenvalues $\lambda_i$. The equation-of-state parameter is then expressed as
\begin{equation}
1+ w(z) = \sum_{i=1}^N \alpha_i\, e_i(z)~,
\label{eq:w_expand}
\end{equation}
where the $e_i(z)$ are the principal components. 
The coefficients $\alpha_i$, which can be computed via the orthonormality
condition
\begin{equation}
\alpha_i = \int (1+w(z)) e_i(z) dz
\end{equation}
are each determined with an accuracy $1/\sqrt{\lambda_i}$. Several of these
components are shown for a future SN survey in the right panel of
Fig.~\ref{fig:w0wp_PC}, while measurements of the first six PCs of the
equation of state from the current (and predictions for future) data are shown
in Fig.~\ref{fig:pc_1d}.  

There are multiple advantages of using the PCs of dark energy (of either the
equation of state $w(z)$, or of $\rhode(z)$ or $H(z)$):

\begin{itemize}
\item The method is as close to ``model independent'' as one can realistically
  get;

\item Data tells us what we measure and how well; there are no arbitrary
  parametrizations imposed;

\item One can use this approach to design a survey that is most sensitive to the
dark energy equation-of-state parameter in some specific redshift interval...

\item ...or to study how many independent parameters are measured well by a
  combination of cosmological probes (i.e.\ how many PCs have
  $\sigma(\alpha_i)$ or $\sigma(\alpha_i)/\alpha_i$ less than some threshold
  value \cite{dePutter_Linder}).
\end{itemize}

There are a variety of useful extensions of this method,
including uncorrelated measurements of the equation-of-state parameters in redshift
intervals \cite{Huterer_Cooray}.

\bigskip
{\bf Figures of Merit.}  We finally discuss the so-called figures of merit
(FoMs) for dark energy experiments. A FoM is a number, or collection of
numbers, that serve as simple and quantifiable metrics by which to evaluate
the accuracy of constraints on dark energy parameters from current and
proposed experiments.  For example, marginalized accuracy in the (constant)
equation of state, $w$, could serve as a figure of merit -- since a large FoM
is ``good'', we could simply define ${\rm FoM} = 1/\sigma_w$, or
$1/\sigma_w^m$ where $m$ is some positive power.

The most commonly discussed figure of merit is that proposed by the Dark
Energy Task Force (\citet{DETF}, though this proposal goes back to
\citet{Hut_Tur_00}), which is essentially inverse area in the $w_0$--$w_a$
plane. For uncorrelated $w_0$ and $w_a$ this would be $\propto
1/(\sigma_{w_0}\times \sigma_{w_a})$; because the two are typically
correlated, the FoM can be defined as
\begin{equation}
{\rm FoM}^{(w_0-w_a)} \equiv (\det {\bf C})^{-1/2} \approx  {6.17 \pi \over A_{95}},
\label{eq:DETF_FoM}
\end{equation}
where ${\bf C}$ is the $2\times 2$ covariance matrix in $(w_0, w_a)$ after
marginalizing over all other parameters, and $A_{95}$ is the area of the 95.4\%~CL
region in the $w_0$--$w_a$ plane.  Note that the constant of proportionality is not
important, since typically we compare the FoM from different surveys, and the
constant disappears when we take the ratio.

\begin{figure}[t]
\begin{center}
\includegraphics[height=3.0in,width=2.8in]{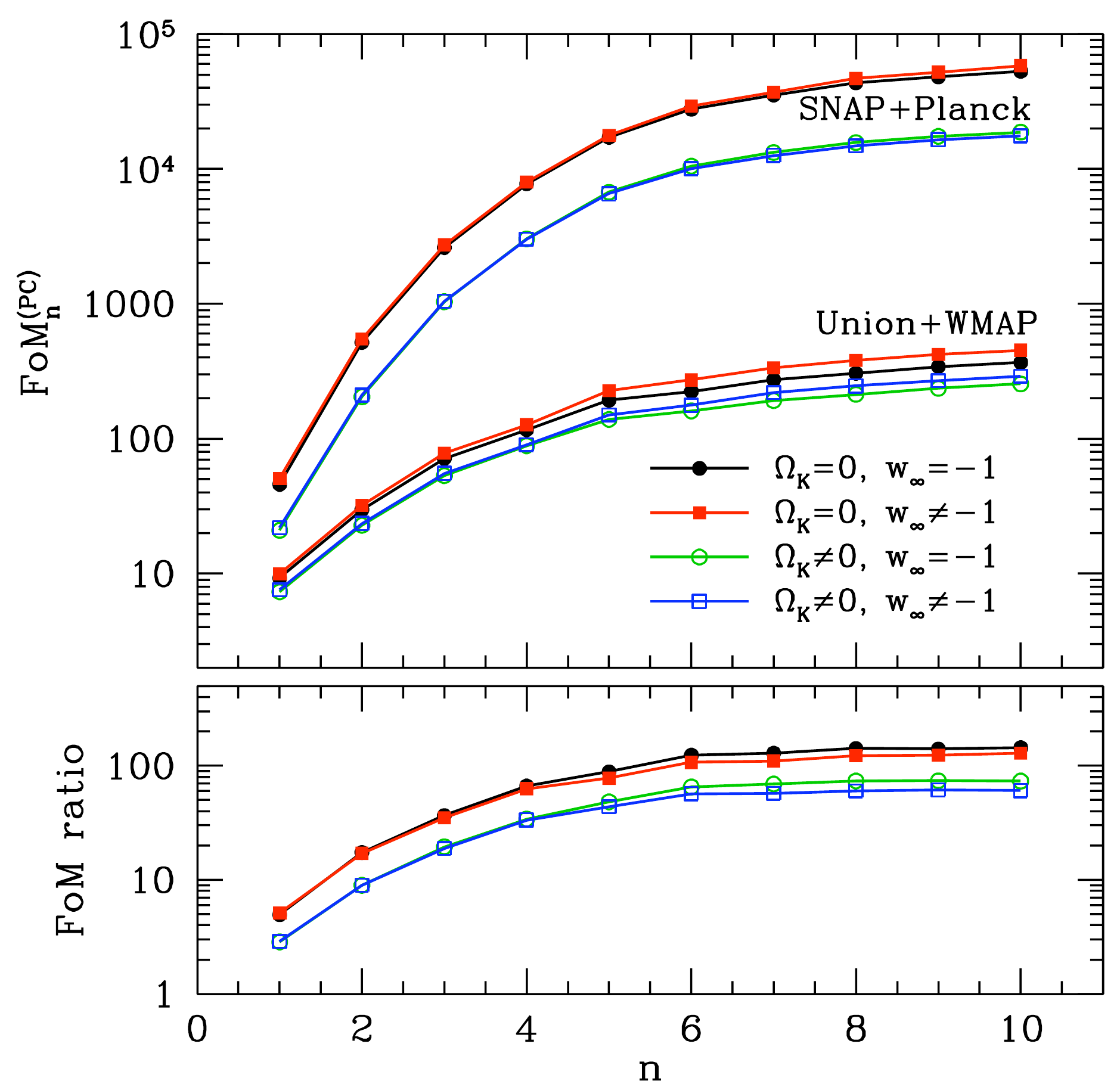}
\end{center}
\caption{ Current and future figures of merit (FoM) for dark energy surveys,
  based on the principal-component (PC) based FoM from
  Eq.~(\ref{eq:fompc1}). {\it Top panel:} PC figures of merit ${\rm
    FoM}_n^{({\rm PC})}$ with forecasted uncertainties  for  a combination
  of planned and ongoing space telescopes SNAP(SN)+Planck and with measured
  uncertainties for already completed surveys Union+WMAP.  {\it Bottom panel:}
  Ratios of ${\rm FoM}_n^{({\rm PC})}$ forecasts to current values.  In both
  panels, point types indicate different quintessence model classes: flat
  (solid points) or non-flat (open points), either with (squares) or without
  (circles) early dark energy. Adopted from \citet{MHH_FoM}. }
\label{fig:fom_pc}
\end{figure}

While the standard ``DETF FoM'' defined in Eq.~(\ref{eq:DETF_FoM}) keeps some
information about the dynamics of DE (that is, the time variation of $w(z)$),
several other general FoMs have been proposed. For example, \citet{MHH_FoM}
proposed taking the FoM to be inversely proportional to the volume of the
n-dimensional ellipsoid in the space of principal component parameters
\begin{equation}
{\rm FoM}^{({\rm PC})}_n \equiv 
\left({ \det {\bf C}_n  \over \det {\bf C}_{n}^{(\rm prior)}}  \right)^{-1/2},
\label{eq:fompc1}
\end{equation}
where the prior covariance matrix is again unimportant since it would cancel
in the comparison of ratios of the FoMs. Fig.~\ref{fig:fom_pc}, near the end
of this Chapter, illustrates this FoM for current and future surveys. 

\section{Other probes of dark energy}\label{sec:other_probes}

In addition to type Ia supernovae, there are several other important probes of
dark energy. These probes operate using very different physics, and have very
different systematic errors. 

The principal probes, in addition to SN Ia, are baryon acoustic
oscillations, weak gravitational lensing, and galaxy cluster abundance.  We
will now discuss each of those in turn.  Additionally, there are secondary
probes of dark energy --- ones that might be useful for DE, but are currently
not as well developed as the primary probes. We will discuss these briefly as
well.

\bigskip
{\bf Baryon acoustic oscillations (BAO).}  BAO refers to the signature of
acoustic oscillations which are imprinted into the present-day correlations of
galaxies by baryonic physics at the epoch of recombination (for a popular
review, see \citet{Eisenstein_NewAstRev}).  Measurements of the length scale
characteristic of these oscillations, roughly $100\hinvmpc$ comoving, enable
inferring the angular diameter distance out to galaxies probed in a survey,
and thus a robust way to measure the energy contents of the universe.

Note that the power spectrum of density perturbations in dark matter, $P(k)$, is
mainly sensitive to the density in matter (relative to critical),
$\Omega_M$. If we assume a flat universe (either motivated by the inflationary
``prior'', or by recent data), then $\ode = 1-\Omega_M$ and measurements of
the broad-band shape of the power spectrum can get the dark energy density,
but not the equation of state $w$.

\begin{figure}[t]
\begin{center}
\includegraphics[height=2.6in,width=2.6in]{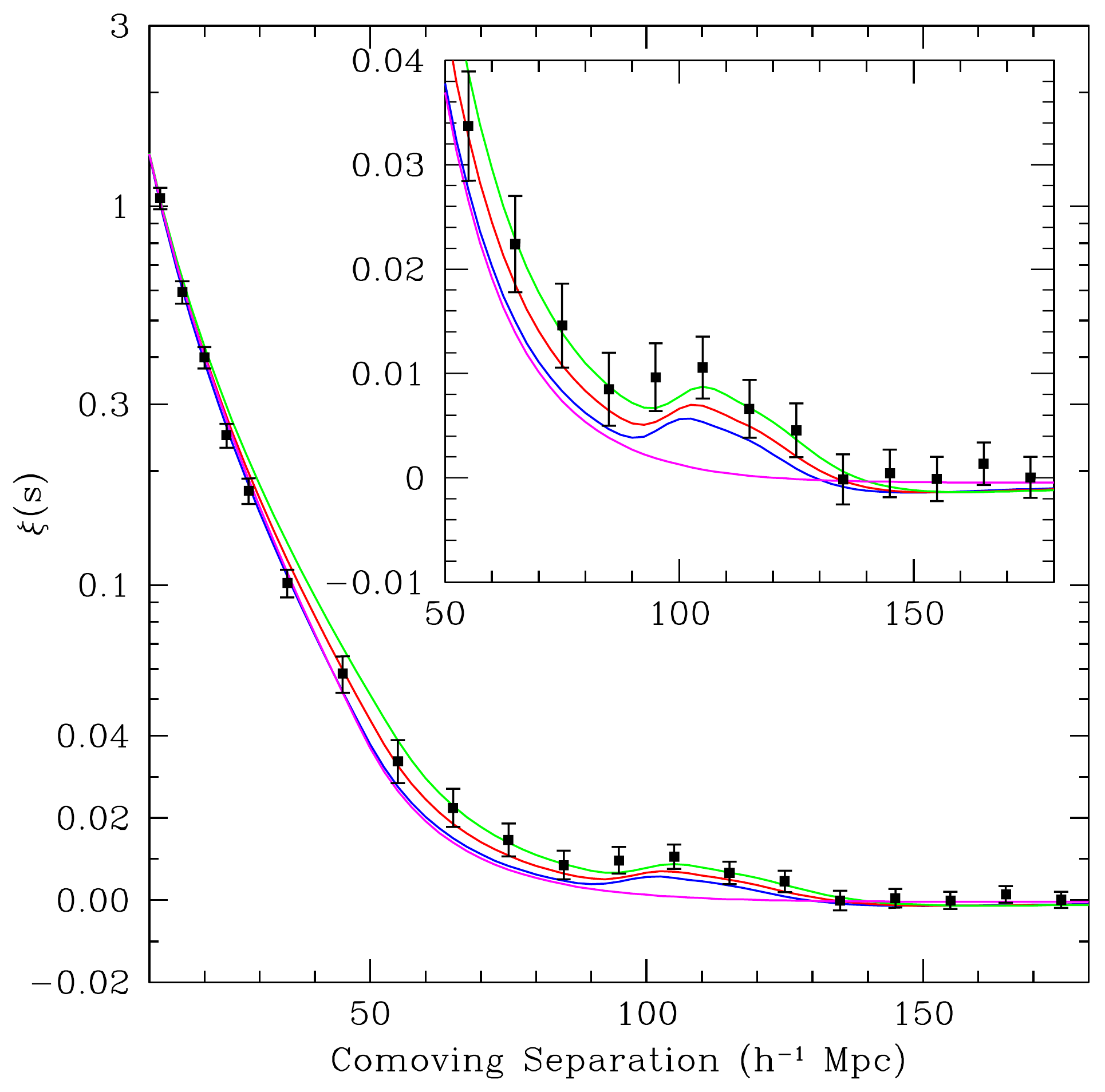}
\end{center}
\caption{Detection of the baryon acoustic peak in the clustering of luminous
  red galaxies in the SDSS \cite{SDSS_BAO}.  Shown is the two-point galaxy
  correlation function in redshift space; inset shows an expanded view with a
  linear vertical axis.  Curves correspond to $\Lambda$CDM predictions for
  $\Omega_{\rm M} h^2=0.12$ (green), $0.13$ (red), and $0.14$ (blue). Magenta
  curve shows a $\Lambda$CDM model without BAO. } 
\label{fig:bao}
\end{figure}

However, the small ($\sim 10\%$) {\it oscillations} in the power spectrum
provide much more information about DE.  The BAO data determine the ratio of
the sound horizon at last scattering to the quantity $D_V(z)\equiv
[z\,r^2(z)/H(z)]^{1/3}$ at the measured redshift; given that the sound horizon
is independently determined rather accurately, the BAO approximately provides
measurement of distance to the redshift where the galaxies reside. For
example, \citet{PercivalBAO} analyze combined data from 2-degree Field Galaxy
Redshift Survey and the Sloan Digital Sky Survey which measure the clustering
at mean redshifts $z=0.2$ and $z=0.35$ respectively.

Key to successful application of baryon acoustic oscillations are redshift
measurements of galaxies in the sample. We need the galaxy redshifts in order
to know where to ``put them'' in three dimensions, and thus to reconstruct the
precise length scale at which the slight excess of clustering occurs. Another
systematic that needs to be understood is the bias of galaxies in the sample
(whose clustering we measure) to the underlying dark matter (whose clustering
we can predict); if the bias has scale-dependent features on scales of $\sim
100\Mpc$, then the systematic errors creep in.  Future surveys that plan to
utilize this method typically propose measuring redshifts of millions of
galaxies, and the goal is to go deep ($z\sim 1$, and beyond) and have wide
angular coverage as well.

Let us finally say a few words about the measured quantity, the power
spectrum. In the dimensionless form, it is given by
\begin{equation}
\Delta^2(k)\equiv {k^3 P(k)\over 2\pi^2} = A\,{4\over 25}
{1\over \Omega_M^2}\left ({k\over k_{\rm piv}}\right )^{n-1}
\left ({k\over H_0}\right )^4 \,D(z)^2\, T^2(k)\, T_{\rm nl}(k)\,,
\label{eq:PS_formula}
\end{equation}
where $A$ is the normalization of the power spectrum (for the concordance
cosmology, $A\simeq 2.4\times 10^{-9}$), $k_{\rm piv}$ is the ``pivot'' around
which we compute the spectral index $n$ ($k_{\rm piv}=0.002\Mpc^{-1}$ is
often used); $D(z)$ is the linear growth of perturbations normalized to unity
today; $T(k)$ is the transfer function that describes evolution of
fluctuations inside the horizon and across the matter-radiation transition epoch and
which encodes the BAOs; $T_{\rm nl}$ is a prescription for the ${\it nonlinear}$
power spectrum which is relevant at small scales (e.g.\ $k\gtrsim
0.2\hmpcinv$ today).  Notice that $\Delta^2\propto k^{n+3}$, and thus
$P(k)\propto k^n$, with $n\simeq 1$, was predicted by Harrison, Zeldovich and
Peebles in the late 1960s; this was a decade before inflation was proposed,
and about three decades before measurements confirmed that $n\simeq 1$!

\begin{figure}[!t]
\centerline{
\includegraphics[height=1.8in,width=2.2in]{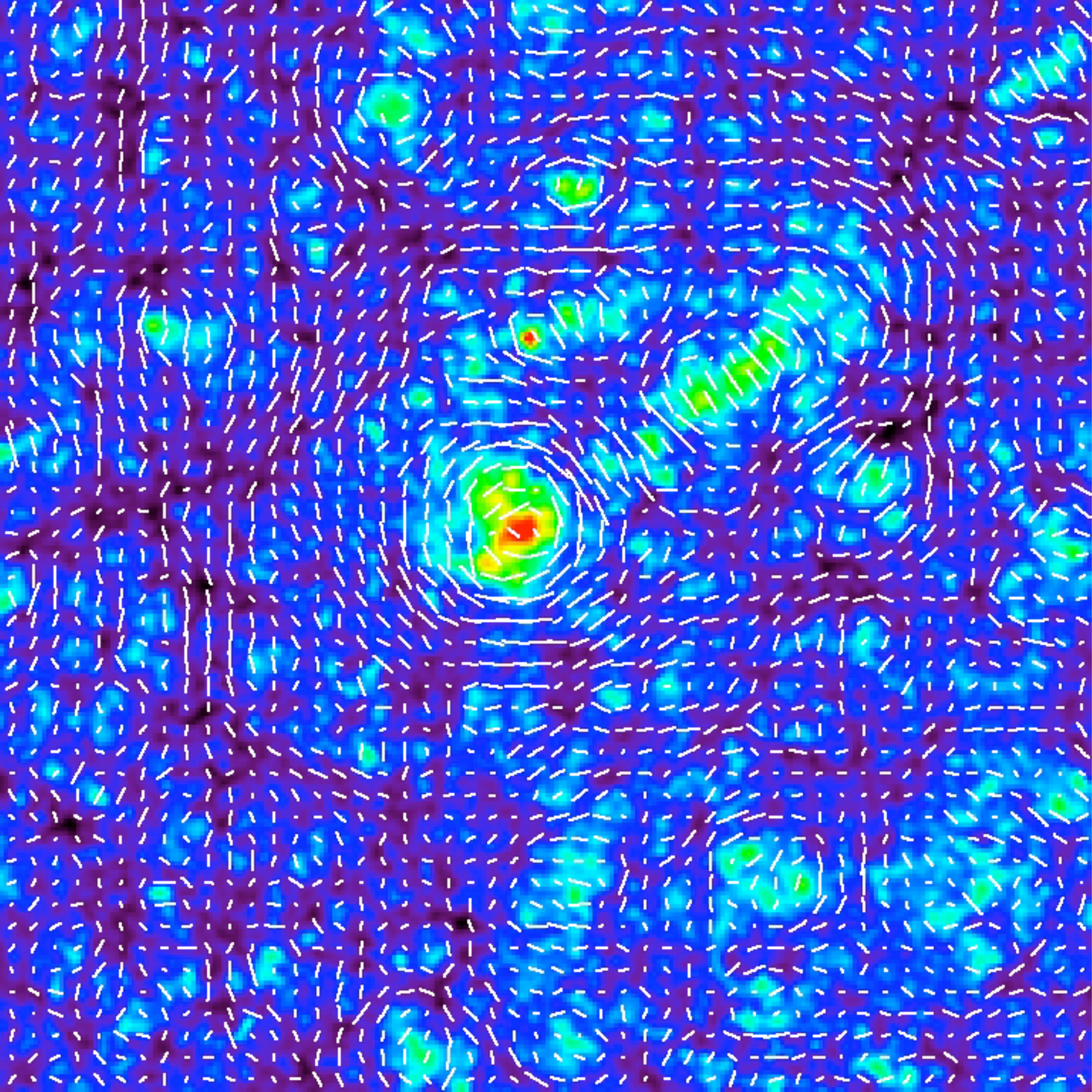}\hspace{0.1cm}
\includegraphics[height=2.0in,width=2.9in]{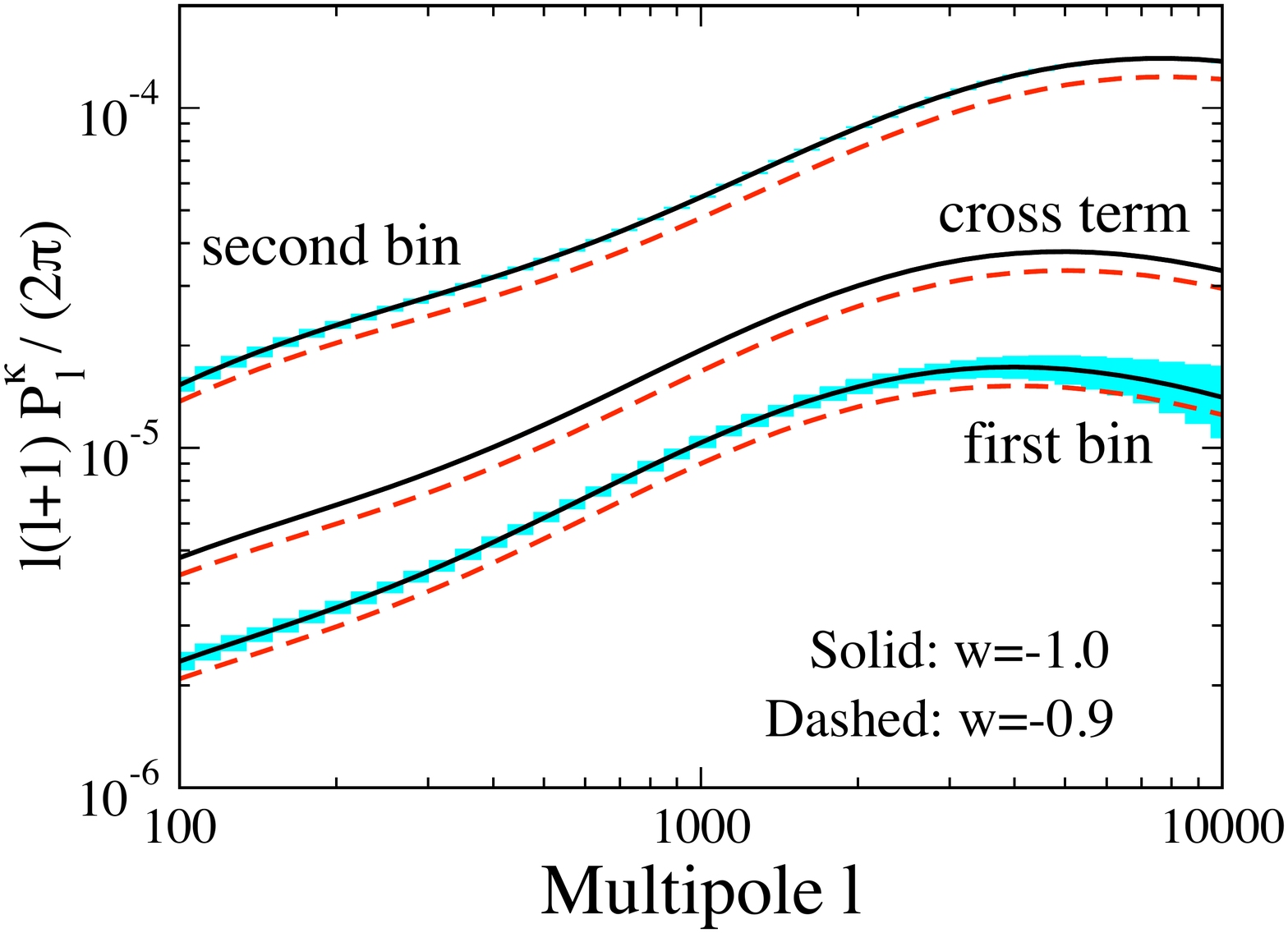}
}
\caption{{\it Left panel:} Cosmic shear field (white ticks) superimposed on
  the projected mass distribution from a cosmological N-body simulation:
  overdense regions are bright, underdense regions are dark. Note how the
  shear field is correlated with the foreground mass distribution.  Figure
  courtesy of T.\ Hamana. {\it Right panel:} Cosmic shear angular power spectrum
  and statistical errors expected for LSST for $w=-1$ and $-0.9$. For
  illustration, results are shown for source galaxies in two broad redshift
  bins, $z_s=0-1$ (first bin) and $z_s=1-3$ (second bin); the cross-power
  spectrum between the two bins (cross term) is shown without the statistical
  errors.  }
\label{fig:WL}
\end{figure}

\bigskip
{\bf Weak gravitational lensing.}  The gravitational bending of light by
structures in the Universe distorts or shears images of distant galaxies;
see the left panel of Fig.~\ref{fig:WL}.  This distortion allows the
distribution of dark matter and its evolution with time to be measured,
thereby probing the influence of dark energy on the growth of structure (for
a detailed review, see e.g.\ \citet{Bartelmann_Schneider}; for 
brief reviews, see \citet{Hoekstra_Jain} and \citet{Huterer_GRG}).

Gravitational lensing produces distortions of images of background
galaxies. These distortions can be described as mapping between the
source plane ($S$) and image plane ($I$) 
\begin{equation}
\delta x_i^S=A_{ij} \delta x_j^I\,,
\end{equation}
where $\delta {\bf x}$ are the displacement vectors in the
two planes and $A$ is the distortion matrix
\begin{equation}
A=
\left ( 
\begin{array}{cc}
	1-\kappa-\gamma_1	&	-\gamma_2   \\
	-\gamma_2		& 1-\kappa+\gamma_1 
\end{array}
 \right ).  
\end{equation}

The deformation is described by the convergence $\kappa$ and complex shear
$(\gamma_1, \gamma_2)$; the total shear is defined as
$|\gamma|=\sqrt{\gamma_1^2+\gamma_2^2}$. We are interested in the weak lensing
limit, where $\kappa$, $|\gamma|\ll 1$.  Magnification can be expressed in terms
of $\kappa$ and $\gamma_{1,2}$ as
\begin{equation}
\mu = {1\over |1-\kappa|^2 - |\gamma|^2}\approx 1+2\kappa+O(\kappa^2,
\gamma^2),
\end{equation}
where the second approximate relation holds in the weak lensing limit. 

We can theoretically predict convergence and shear, given a sample of sources
with known redshift distribution and cosmological parameter values.  The
convergence in any particular direction on the sky ${\bf\hat n}$ is given by
the integral along the line-of-sight
\begin{equation}
\kappa({\bf \hat n}, \chi)=\int_0^{\chi} W(\chi')\, 
\delta(\chi')\, d\chi'\, ,
\label{eq:conv}
\end{equation}
where $\delta$ is the perturbation in matter energy density and $
W(\chi)$ is the geometric weight function describing the lensing efficiency of
foreground galaxies.  The most efficient lenses lie about half-way between us
and the source galaxies whose shapes we measure.

The statistical signal due to gravitational lensing by large-scale structure
is termed ``cosmic shear.''  The cosmic shear field at a point in the sky is
estimated by locally averaging the shapes of large numbers of distant
galaxies.  The primary statistical measure of the cosmic shear is the shear
angular power spectrum measured as a function of source galaxy redshift $z_s$.
(Additional information is obtained by measuring the correlations between
shears at different redshifts or with foreground lensing galaxies.)  

The convergence can be transformed into multipole space $\kappa_{lm}=\int d
{\bf\hat n} \,\kappa({\bf \hat n}, \chi)\, Y_{lm}^*({\bf\hat n})$, and the
power spectrum is defined as the two-point correlation function (of
convergence, in this case) $\langle \kappa_{\ell m}\kappa_{\ell'm'}\rangle =
\delta_{\ell \ell'}\, \delta_{m m'} \,P_\ell^{\kappa}$.  The angular power
spectrum is
\begin{equation}
P^\gamma_{\ell}(z_s)\simeq P^\kappa_\ell(z_s)
 = \int_{0}^{z_s}  {dz\over H(z) d_A^2(z)} W(z)^2
P\left (k={\ell \over d_A(z)}; z\right ),
\label{eqn:Limber}
\end{equation}
where $\ell$ denotes the angular multipole, $d_A(z) = (1+z)^{-2}d_L(z)$ is the
angular diameter distance, the weight function $W(z)$ is
the efficiency for lensing a population of source galaxies and is determined
by the distance distributions of the source and lens galaxies, and
$P(k,z)$ is the usual power spectrum of density perturbations. Notice the
  integral along the line of sight:  essentially, weak lensing projects the
density fluctuations between us and the galaxies whose shear we measure. 

The dark-energy sensitivity of the shear angular power
spectrum comes from two factors: 
\begin{itemize}
\item {\it geometry} -- the Hubble parameter, the angular-diameter distance,
  and the weight function $W(z)$; and
\item {\it growth of structure} -- through the redshift evolution of the power
  spectrum $P(k)$ (or more precisely, from the function $D(z)$ in
  Eq.~(\ref{eq:PS_formula})).
\end{itemize}
The {\it three}-point correlation function of cosmic shear is also sensitive
to dark energy, and provides important complementary information about dark
energy (e.g.~\citet{Takada_Jain}).

The  statistical uncertainty in measuring the shear power spectrum on
large scales is 
\begin{equation}
\Delta P^\gamma_\ell = \sqrt{\frac{2}{(2\ell+1)f_{\rm sky}} }
\left[ P^\gamma_\ell +\frac{\sigma^2(\gamma_i)}{n_{\rm eff}} \right]~~,
\label{eqn:power_error}
\end{equation}
where $f_{\rm sky}$ is the fraction of sky area covered by the survey
($\fsky=0.5$ for half-sky, etc), $\sigma^2(\gamma_i)$ is the variance in a
single component of the (two-component) shear (this number is $\sim 0.2$ for
typical measurements), and $n_{\rm eff}$ is the effective number density per
steradian of galaxies with well-measured shapes.  The first term in brackets
dominates on large scales, and comes from sample variance (also known as {\it
  cosmic variance}) due to the fact that only a finite number of samples of
structures are available in our universe.  The second term dominates on small
scales, and represents the shot-noise from the variance in galaxy
ellipticities (``shape noise'') combined with a finite number of galaxies,
hence the inverse proportionality to $n_{\rm eff}$.

The principal systematic errors in weak lensing measurements come from the
limitations in measuring galaxy shapes accurately. There are also systematic
uncertainties due to limited knowledge of the redshifts of source galaxies:
because taking spectroscopic redshifts of most source galaxies will be
impossible (they number in many millions), one has to rely to approximate
photometric redshift techniques, where one gets redshift information from
multiple-wavelength (i.e.\ multi-color) observations.

The right panel of Fig.~\ref{fig:WL} shows the dependence on the dark energy
of the shear power spectrum and an indication of the statistical errors
expected for a survey such as LSST, assuming a survey area of 15,000
sq.\ deg.\ and effective source galaxy density of $n_{\rm eff}=30$ galaxies
per sq.\ arcmin, and divided into two radial slices. Current surveys cover
a more modest $\sim 100$ square degrees, with a comparable or slightly lower
galaxy density. Note that the proportionality of errors to $\fsky^{-1/2}$
means that large sky coverage is at a premium.

\bigskip
{\bf Clusters of galaxies.}  Galaxy clusters are the largest virialized
objects in the Universe.  Therefore, not only can they be observed, but also
their number density can be {\it predicted} quite reliably, both analytically
and from numerical simulations.  Comparing these predictions to measurements
from the large-area cluster surveys that extend to high redshift ($z \gtrsim
1$) can provide precise constraints on the cosmic expansion history.

\begin{figure}[!t]
\centerline{
\includegraphics[height=2.5in,width=2.5in]{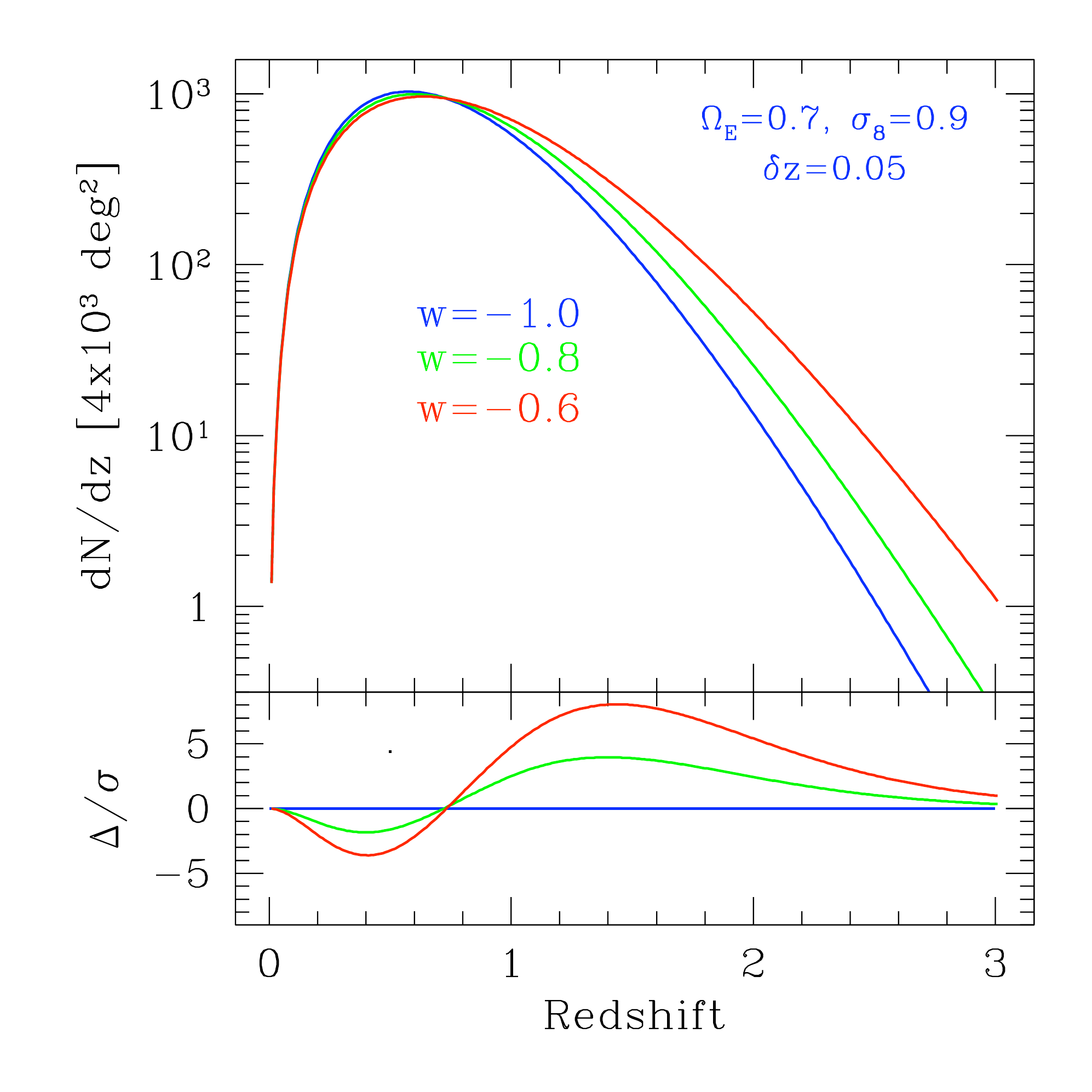}
\includegraphics[height=2.3in,width=2.5in]{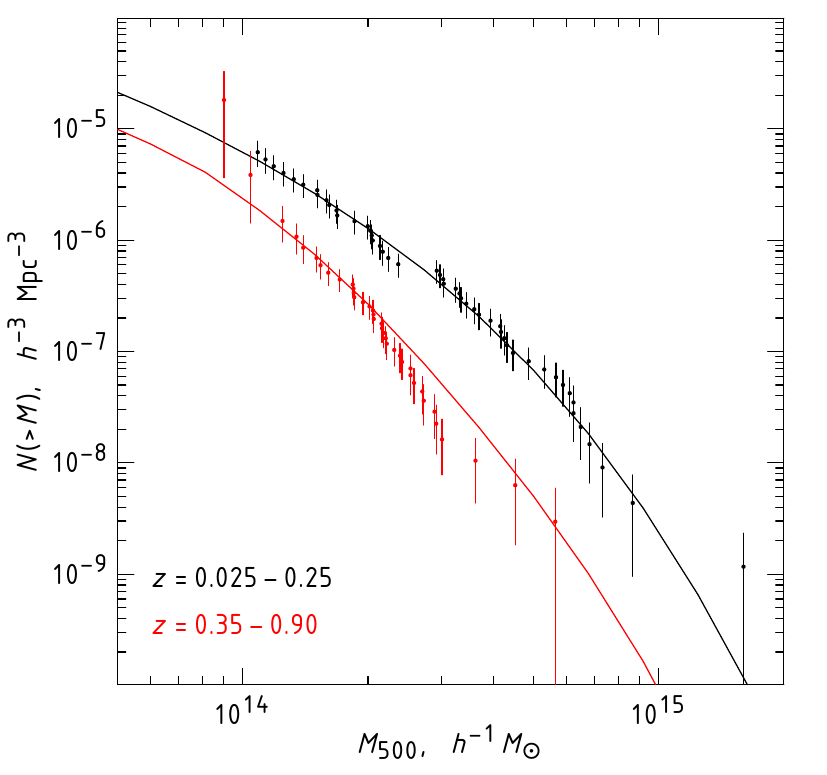}
}
\caption{{\it Left panel:} Predicted cluster counts for a survey covering
  4,000 sq.\ deg.\ that is sensitive to halos more massive than $2\times
  10^{14} M_\odot$, for 3 flat cosmological models with fixed $\Omega_{\rm
    M}=0.3$ and $\sigma_8=0.9$.  Lower panel shows fractional differences
  between the models in terms of the estimated Poisson errors.  From
  \citet{Mohr_04}. {\it Right panel:} Measured mass function -- $n(z, M_{\rm
    min}(z))$ -- in our notation -- from the 400 square degree survey of ROSAT
  clusters followed up by Chandra.  Adopted from \citet{Vikhlinin:2008ym}.  }
\label{fig:massfun}
\end{figure}

The absolute number of clusters in a survey of solid angle $\Omega_{\rm survey}$
centered at redshift $z$ and in the shell of thickness $\Delta z$ is given by
\begin{equation}
N(z, \Delta z)=\Omega_{\rm survey}
\int_{z-\Delta z/2}^{z+\Delta z/2} n(z, M_{\rm min}(z)) 
\,{dV(z)\over d\Omega\,dz}\, dz,
\label{eq:clustercount}
\end{equation}
where $M_{\rm min}$ is the minimal mass of clusters in the survey (usually of
order $10^{14}\Msun$). Note that knowledge of the minimal mass is extremely
important, since the  mass function  $n(z, M_{\rm min}(z))$ is exponentially
decreasing with $M$, so that most of the contribution comes from a small range
of masses just above $M_{\rm min}$. The mass function is key to theoretical
predictions, and it is usually obtained from a combination of analytic and
numerical results; the original mass function used in cosmology is the 36-year
old Press-Schechter mass function \cite{Press_Schechter}, and the more recent work
provides fitting functions to simulations' results that are accurate to
several percent \cite{Tinker}.  Furthermore, the volume element can easily be
related to comoving distance $r(z)$ and the expansion rate $H(z)$ via
$dV(z)/(d\Omega\,dz)= r^2(z)/H(z)$, and it is known exactly for a given
cosmological model.

The sensitivity of cluster counts to dark energy arises -- as in the case of
weak lensing -- from two factors:
\begin{itemize}
\item {\it geometry}, the term $dV(z)/(d\Omega\,dz)$ in Eq.~(\ref{eq:clustercount})
is the comoving volume element 

\item {\it growth of structure}, the mass function $n(z, M_{\rm min}(z))$ depends on the
  evolution of density perturbations.
\end{itemize}
The mass function's near-exponential dependence upon the power spectrum is at
the root of the power of clusters to probe dark energy. More specifically, the
mass function explicitly depends on the {\it amplitude of mass fluctuations}
smoothed on some scale $R$
\begin{equation}
\sigma^2(R, z) = \int_0^\infty \Delta^2(k, z)
\left ({3j_1(kR)\over kR}\right )^2 d\ln k 
\end{equation}
where $\Delta^2(k, z)$ is the dimensionless power spectrum defined in
Eq.~(\ref{eq:PS_formula}), while $R$ is traditionally taken to be $\sim
8\hinvmpc$ at $z=0$ and roughly corresponds to the typical size of a
galaxy cluster. The term in angular parentheses is the Fourier transform of
the top-hat window that averages out the perturbations over regions of radius
$R$.

Systematic errors in cluster counts mainly concern uncertainty in how to
convert from an observable quantity (X-ray light, gravitational lensing
signal, etc) to the mass of a cluster. Current best estimates of mass are at
the level of several  tens of percent per cluster, and there is ongoing effort
to find observable quantities, or combinations thereof, that are tightly correlated with mass. 

The left panel of Fig.~\ref{fig:massfun} shows the sensitivity to the dark
energy equation-of-state parameter of the expected cluster counts for the
South Pole Telescope and the Dark Energy Survey. At low to intermediate
redshift, $z<0.6$, the differences are dominated by the volume element; at
higher redshift, the counts are most sensitive to the growth rate of
perturbations. The right panel shows measurements of the mass function using
recent X-ray observations of clusters.

\bigskip
{\bf Summary of principal probes.} Figure \ref{fig:kowalski} adopted from
\citet{Union2}, summarizes constraints in the $\Omega_M$-$\Omega_\Lambda$ and
$\Omega_M$-$w$ planes (the latter assuming a flat universe) from CMB, BAO and
SN Ia.  In Table 1.1 we list the principal strengths and
weaknesses of the four principal probes of DE.  Control of systematic errors
--- observational, instrumental and theoretical --- is crucial for these
probes to realize their intrinsic power in constraining dark energy.

\begin{table}[!t]
\begin{center}
    \tbl{Comparison of dark energy probes, adopted from
      \citet{Frieman:2008sn}.  CDM refers to Cold Dark Matter paradigm, FoM is
      the Figure-of-Merit for dark energy surveys defined in the Dark Energy
      Task Force (DETF) report, while SZ refers to Sunyaev-Zeldovich effect.
    }
%
%
{
  \begin{tabular}{llll} \hline\hline
\rule[-2mm]{0mm}{6mm}Method & Strengths & Weaknesses & Systematics \\ \hline
WL & growth+geometry, & CDM assumptions & Shear systematics,    \\[-0.5ex]
& Large FoM & & Photo-z                    \\\hline
SN & pure geometry, & complex physics & evolution, \\[-0.5ex]
& mature &  & dust extinction              \\\hline
BAO & pure geometry, & coarse-grained & bias, non-linearity,   \\[-0.5ex]
& low systematics & information & redshift distortions         \\\hline
CL & growth+geometry,  & CDM assumptions & mass-observable,     \\[-0.5ex]
& X-ray+SZ+optical &  & selection function   \\\hline
\hline
    \end{tabular}
}
\end{center}
    \label{tab:probe_sys}
\end{table}

\bigskip
{\bf Role of the CMB.}  While the CMB provides precise cosmological
constraints, by itself it has little power to probe dark energy.  The reason
is simple: the CMB provides a single snapshot of the Universe at a time when
dark energy contributed a tiny part of the total energy density (a part in
$10^9$ if dark energy is the vacuum energy, or when $w=-1$).  Nevertheless,
the CMB plays a critical supporting role by determining other cosmological
parameters, such as the spatial curvature and matter density, to high
precision, thereby considerably strengthening the power of the methods
discussed above. Essentially, what we get from the CMB is a {\it single}
measurement of the angular diameter distance to recombination, $d_A(z\approx
1000)$ -- therefore it provides a {single} very accurate measurement of the
parameters $\Omega_M$, $\ode$ (if we do not assume a flat universe), and $w$
(or $w(z)$ if we don't assume that the equation of state is constant). So,
while the CMB alone suffers from degeneracy between the DE parameters, it is
indispensable in breaking parameter degeneracies present in other cosmological
probes; see \citet{frieman_03} for more details.  Data from the Planck CMB
mission, launched in 2009, will therefore strongly complement those from dark energy surveys.

\bigskip
{\bf Secondary probes.} There are a number of secondary probes of dark
energy; here we review some of them. 

\begin{itemize}
%
\item The Integrated Sachs-Wolfe (ISW) effect provided a confirmation of
  cosmic acceleration. ISW impacts the large-angle structure of the CMB
  anisotropy, but low-$\ell$ multipoles are subject to large cosmic variance,
  limiting the power of this probe.  Nevertheless, ISW is of interest because it is
  able to reveal the imprint of large-scale dark-energy perturbations
  \cite{Hu_Scranton}.

\item Gravitational radiation from inspiraling binary neutron stars or black
  holes can, if detected in the future, serve as ``standard sirens'' to
  measure absolute distances \cite{Holz_Hughes_05}. If their redshifts can be
  determined, then they could be used to probe dark energy through the Hubble
  diagram \cite{Dalal06}.

\item Long-duration gamma-ray bursts have been proposed as standardizable
  candles \cite{schaefer03}, but their utility as cosmological distance
  indicators that could be competitive with or complementary to SN Ia has yet
  to be established. 

\item The optical depth for strong gravitational lensing (multiple imaging) of
  QSOs or radio sources has been proposed and used to provide independent
  evidence for dark energy, though these measurements depend on modeling the
  density profiles of lens galaxies.

\item The redshift drift effect (also known as the Sandage-Loeb effect
  \cite{Sandage62,Loeb}) -- the redshift change of an object measured using
  extremely high-resolution spectroscopy over a period of $10$ years or more
  -- may some day be useful in constraining the expansion history at higher
  redshift, $2\lesssim z\lesssim 5$ \cite{Cor_Hut_Mel}.

\item Polarization measurements from distant galaxy clusters -- which probe
  the quadrupole of the CMB radiation at the epoch when the cluster light was
  emitted, and therefore the gravitational potential at that epoch -- in
  principle provide a sensitive probe of the growth function and hence dark
  energy \cite{Coo_Hut_Bau}.

\item The relative ages of galaxies at different redshifts, if they can be
  determined reliably, provide a measurement of $dz/dt$ and, from
  \begin{equation}
  t(z) = \int_0^{t(z)} dt' = \int_z^\infty {dz' \over (1+z')H(z')},
  \label{eq:cosmic_time}
  \end{equation}
  measure the expansion history directly \cite{Jimenez_Loeb}. 
\end{itemize}

\section{The accelerating universe:  summary} 

There are the five important things to know about dark energy:

\begin{enumerate}
\item Dark energy has negative pressure. It can be described with its energy
  density relative to critical today $\ode$, and equation of state $w\equiv
  p_{\rm DE}/\rho_{\rm DE}$; the cosmological constant (or vacuum energy) has
  $w=-1$ precisely and at all times. More general explanations for dark energy
  may have constant or time dependent equation of state. 
   Assuming constant $w$, current constraints roughly give $w\approx
  -1\pm 0.1$. Measuring the equation of state (and its time dependence) may
  help understand the nature of dark energy, and is a key goal of modern
  cosmology. 
\item The accelerating universe quenches gravitational collapse of large
  structures and suppresses the growth of density perturbation: whenever dark
  energy dominates, structures do not grow, essentially because the expansion
  is too rapid. 
\item Dark energy comes to dominate the density of the universe only recently,
  at $z\lesssim 1$. At earlier epochs, dark energy density is small
  relative to matter density. 
\item Dark energy is spatially smooth. It affects both the geometry (that is,
  distances in the universe) and the growth of structure  (that is, clustering
  and abundance of galaxies and clusters of galaxies). 
\item Dark energy can be probed using a variety of cosmological probes that
  measure geometry (i.e.\ the expansion history of the universe) and the growth of
  structure. Control of systematic errors in these cosmological probes is key
  to their success in measuring the properties of dark energy.
\end{enumerate}

\section*{Acknowledgments} 

I thank  Bob Kirshner, Eric Linder and Adam Riess for many useful comments on
an earlier version of this manuscript, and to Josh Frieman and Michael Turner
for collaboration on an earlier review \cite{Frieman:2008sn} that helped me
organize thoughts about dark energy. I am supported by DOE OJI grant under
contract DE-FG02-95ER40899, NSF under contract AST-0807564, and NASA under
contract NNX09AC89G.

\bibliographystyle{ws-book-har}    
\bibliography{bibfile}      

\end{document}